\documentclass[12pt, draftclsnofoot, onecolumn]{IEEEtran}

\usepackage{amsmath,amsfonts,mathtools,amsthm,amssymb,dsfont,multirow,hhline,float,color}
\usepackage{graphicx}
\usepackage{algorithmic,algorithm}
\usepackage{amsthm}
\usepackage{subfigure}
\usepackage{epstopdf}
\usepackage[noadjust]{cite}

\usepackage[flushleft]{threeparttable}

\newtheorem{lemma}{Lemma}
\newtheorem{prop}{Proposition}

\newcommand{\tr}{\mathrm{tr}}
\newcommand{\vect}{{\mathrm{vec}}}

\newcommand{\Fopt}{{\mathbf{F}_\mathrm{opt}}}
\newcommand{\FRF}{{\mathbf{F}_\mathrm{RF}}}
\newcommand{\fopt}{{\mathbf{f}_\mathrm{opt}}}
\newcommand{\fRF}{{\mathbf{f}_\mathrm{RF}}}
\newcommand{\FBB}{{\mathbf{F}_\mathrm{BB}}}
\newcommand{\FBD}{{\mathbf{F}_\mathrm{BD}}}
\newcommand{\WBB}{{\mathbf{W}_\mathrm{BB}}}
\newcommand{\WRF}{{\mathbf{W}_\mathrm{RF}}}
\newcommand{\NRFr}{{N_\mathrm{RF}^\mathrm{r}}}
\newcommand{\NRFt}{{N_\mathrm{RF}^\mathrm{t}}}
\newcommand{\Nt}{{N_\mathrm{t}}}
\newcommand{\Nr}{{N_\mathrm{r}}}

\ifCLASSINFOpdf
\else
\fi

\begin{document}
	\title{Doubling Phase Shifters for Efficient Hybrid Precoder Design in Millimeter-Wave Communication Systems}
	\author{Xianghao~Yu,~\IEEEmembership{Student Member,~IEEE},
		Jun~Zhang,~\IEEEmembership{Senior Member,~IEEE},
		and~Khaled~B.~Letaief,~\IEEEmembership{Fellow,~IEEE}
		\thanks{This work was presented in part at Asilomar Conference on Signals, Systems, and Computers, Pacific Grove, CA, USA, Nov. 2016 \cite{asilomar}.}
		\thanks{X. Yu, J. Zhang, and K. B. Letaief are with the Department of Electronic and Computer Engineering, the Hong Kong University of Science and Technology (HKUST), Kowloon, Hong Kong (e-mail: \{xyuam, eejzhang, eekhaled\}@ust.hk).
			K. B. Letaief is also with Hamad Bin Khalifa University, Doha, Qatar (e-mail: kletaief@hbku.edu.qa).
			
			This work was supported by the Hong Kong Research Grants Council under Grant No. 16210216. 
		}
	}
	
	\maketitle
	
	\begin{abstract}
		Hybrid precoding is a cost-effective approach to support directional transmissions for millimeter-wave (mm-wave) communications, but its precoder design is highly complicated. In this paper, we propose a new hybrid precoder implementation, namely the \emph{double phase shifter} (DPS) implementation, which enables highly tractable hybrid precoder design. Efficient algorithms are then developed for two popular hybrid precoder structures, i.e., the fully- and partially-connected structures. For the fully-connected one, the RF-only precoding and hybrid precoding problems are formulated as a least absolute shrinkage and selection operator (LASSO) problem and a low-rank matrix approximation problem, respectively. In this way, computationally efficient algorithms are provided to approach the performance of the fully digital one with a small number of radio frequency (RF) chains. 
		On the other hand, the hybrid precoder design in the partially-connected structure is identified as an eigenvalue problem. To enhance the performance of this cost-effective structure, dynamic mapping from RF chains to antennas is further proposed, for which a greedy algorithm and a modified K-means algorithm are developed. Simulation results demonstrate the performance gains of the proposed hybrid precoding algorithms over existing ones. It shows that, with the proposed DPS implementation, the fully-connected structure enjoys both satisfactory performance and low design complexity while the partially-connected one serves as an economic solution with low hardware complexity. 
	\end{abstract}
	
	
	
	\IEEEpeerreviewmaketitle
	
	\section{Introduction}
	The proliferation of smart mobile devices has resulted in an ever-increasing wireless data explosion, which calls for an exponential increase in the capacity of wireless networks. In particular, the upcoming 5G networks require a 1000X increase in capacity by 2020 \cite{6824752}. 
	The spectrum crunch in current wireless systems stimulates extensive interests on exploiting new spectrum bands for cellular communications, and millimeter-wave (mm-wave) bands from 30 GHz to 300 GHz have been demonstrated to be promising candidates in recent experiments \cite{6515173}.
	Thanks to the smaller wavelength of mm-wave signals, large-scale antenna arrays can be leveraged at both the transmitter and receiver sides, which can provide spatial multiplexing gains with the help of multiple-input multiple-output (MIMO) techniques.
	On the other hand, the ten-fold increase of the carrier frequency introduces several challenges to mm-wave communication systems, especially the high power consumption and cost of hardware components at mm-wave bands \cite{rappaport2014millimeter}. In addition, the large available bandwidth at mm-wave frequencies induces severe frequency selectivity, for which multicarrier techniques such as orthogonal frequency-division multiplexing (OFDM) shall be utilized.
	All the above-mentioned design aspects should be taken into consideration when developing practical transceivers for mm-wave MIMO systems.
	
	By utilizing a small number of radio frequency (RF) chains to combine a low-dimensional digital baseband precoder and a high-dimensional analog RF precoder, hybrid precoding stands out as a cost-effective transceiver solution for mm-wave MIMO systems \cite{1519678,6717211,7579557}.
	Compared with conventional MIMO systems, the additional high-dimensional analog RF precoder is the differentiating part.
	According to the mapping strategies from RF chains to antennas in the analog RF precoder, hybrid precoders can be categorized into the fully- and partially-connected structures \cite{7397861}. In the fully-connected structure, each antenna is connected to all the RF chains. In contrast, each antenna is connected to one RF chain in the partially-connected structure, with a significant reduction in the hardware complexity.
	
	To effectively reduce the power consumption in the RF domain, analog RF precoders are usually implemented by phase shifters at the expense of sacrificing the ability to adjust the amplitude of the RF signals \cite{1519678}. Thus, the analog component forms the major challenge in designing hybrid precoders. 
	Given the large dimension of the design space and the unit modulus constraint induced by the phase shifter implementation, an important design aspect of hybrid precoders is the computational complexity. 
	While various attempts have been made to balance the performance and computational complexity, there is no systematic approach to design \emph{computationally efficient hybrid precoders} with satisfactory performance in the meanwhile. In this paper, we will show the great potential to develop efficient hybrid precoding algorithms by adopting a novel double phase shifter (DPS) hybrid precoder implementation.
	
	\subsection{Related Works and Motivation}
	Most existing works on hybrid precoding focused on the fully-connected structure \cite{6717211,7397861,7389996,6874567,6884253,7448873,7335586,7037444,6928432,7913599,zhang2014achieving,7387790}. The initial efforts started from single-user single-carrier\footnote{In this paper, single-carrier systems refer to single-carrier transmissions over flat-fading channels.} mm-wave systems \cite{6717211,7397861,7389996,6874567}. Then, the investigation was extended to single-user multicarrier \cite{7397861,6884253,7448873} and multiuser single-carrier systems \cite{7335586,7037444,6928432,7913599}. The main differences in these existing works are the approaches in dealing with the unit modulus constraints on the analog RF precoder. 
	
	By choosing the analog beamforming vectors from a predefined candidate set, e.g., array response vectors in \cite{6717211,6884253,7335586,7037444} and discrete Fourier transform beamformers in \cite{6874567}, a greedy algorithm called orthogonal matching pursuit (OMP) has been widely used in designing hybrid precoders. Although its computational complexity is relatively low, the performance is not satisfactory and has been improved by several followed-up works. In \cite{7397861}, it was shown that the unit modulus constraints define a Riemannian manifold, and manifold optimization was introduced to directly tackle them, which helps to approach the performance of the fully digital one with a small number of RF chains. Furthermore, the contribution of each phase shifter to the spectral efficiency was identified in \cite{7389996,7913599}, based on which the analog precoder was optimized in a phase shifter-by-phase shifter fashion. However, these algorithms all involve iterative procedures to optimize the analog RF precoders, which results in high computational complexity. Moreover, there were also some studies on how to achieve the performance of the fully digital precoder with the hybrid structure \cite{zhang2014achieving,7387790}, yet requiring a large number of RF chains, which, to some extend, deviates from the motivation of hybrid precoding.
	
	On the other hand, less attention has been paid on hybrid precoding in the partially-connected structure.
	In \cite{7006720,6824962}, codebook-based design of hybrid precoders was presented for single-user narrowband and OFDM systems, respectively. While using codebook enjoys a low complexity, there will be certain performance loss, and how to design the codebook remains to be clarified. By migrating the concept of successive interference cancellation, an iterative hybrid precoding algorithm in the partially-connected structure was proposed in \cite{7445130} for single-user single-carrier systems. Since the partially-connected structure employs much fewer phase shifters, there should be some inevitable degradation in the analog precoding gain, which makes it difficult for such structure to achieve a high spectral efficiency, especially when the analog precoder is shared across all the users and subcarriers as in the multiuser multicarrier systems. Hence, how to efficiently use the limited number of phase shifters is an urgent issue to be solved in the partially-connected structure.
	
	As illustrated above, in both the fully- and partially-connected structures, there is no comprehensive way to efficiently design hybrid precoders with satisfactory performance, which motivates us to seek a new hybrid precoding architecture that can relieve us from the current dilemma. Furthermore, it is still unclear how to design hybrid precoders in multiuser multicarrier systems, where a single analog RF precoder is shared by a large number of subcarriers, and multiple users that interfere with each other. In this paper, we propose a novel DPS implementation for hybrid precoding in the general setting of multiuser OFDM mm-wave MIMO systems. 
	Although similar implementations were considered in \cite{7387790,7802579},
	the systematic design approach and algorithmic advantages of this new implementation have not been exploited, which will be illustrated in this paper via effective algorithms for different hybrid precoder structures.
	
	\subsection{Contributions}
	{Conventionally, a single phase shifter is used to connect an RF chain and an antenna, i.e., the SPS implementation, which introduces the unit modulus constraints and hinders efficient algorithm design. In this paper, to overcome this algorithmic difficulty, we propose a novel hybrid precoder implementation that makes the precoder design more tractable. Our main contributions are summarized as follows.}
	\begin{itemize}
		\item We propose a novel hybrid precoder implementation, i.e., the DPS implementation, which relaxes the unit modulus constraints of the analog RF precoder and thus {enables computationally efficient} hybrid precoder design. To the best of the authors' knowledge, this is the first attempt to directly adopt the DPS implementation for designing hybrid precoders in multiuser OFDM mm-wave MIMO systems.
		\item For the fully-connected structure, the optimization of the analog RF precoder is formulated as a least absolute shrinkage and selection operator (LASSO) problem, based on which efficient algorithms are developed. Furthermore, the hybrid precoder design is identified as a low-rank matrix approximation problem, which has a closed-form solution. {\color{black}Furthermore, the efficient algorithm for the DPS implementation inspires an effective heuristic hybrid precoder design for the conventional SPS implementation, which outperforms the state-of-the-art algorithms in both computational complexity and spectral efficiency.} 
		\item For the partially-connected structure, we identify that the hybrid precoder design is an eigenvalue problem, and provide closed-form solutions for both analog RF and digital baseband precoders. To further improve the system performance, a dynamic partially-connected structure is proposed. Two effective algorithms, i.e., the greedy and modified K-means algorithms, are proposed to dynamically optimize the mapping strategies from RF chains to antennas.
		\item For both structures, we discover that the hybrid precoder in the multiuser setting will produce residual interuser interference, as it only approximates the fully digital precoder. To this end, we propose to cascade an additional block diagonalization (BD) precoder at the baseband to cancel the interuser interference, which is shown to be effective to further improve the spectral efficiency and multiplexing gain.
		\item Analytical results on the performance gap between the fully- and partially-connected structures are provided. Furthermore, extensive comparisons are offered via simulations to unravel valuable design insights. In particular, the proposed algorithm helps the fully-connected structure to easily approach the performance of the fully digital precoder with a reasonably small amount of RF chains, which cannot be achieved by the widely used OMP algorithm. On the other hand, for the partially-connected structure, it turns out that the dynamic mapping from RF chains to antennas is crucial to achieve good performance. Furthermore, while the DPS partially-connected structure employs much fewer phase shifters, its performance is comparable to the SPS fully-connected structure with the OMP algorithm, which shows its great potential for practical implementation.
	\end{itemize}
	
	\subsection{Organization}
	The remainder of this paper is organized as follows. We introduce the system model and the problem formulation in Section \ref{SecII}. Then, hybrid precoder design for the fully- and partially-connected structures are demonstrated in Section \ref{SecIII} and Section \ref{SecIV}, respectively. Simulation results will be presented in Section \ref{SecV}. Finally, we conclude this paper in Section \ref{SecVI}.
	
	\subsection{Notations}
	The following notations are used throughout this paper. The imaginary unit is denoted as $\jmath=\sqrt{-1}$; $\mathbf{a}$ and $\mathbf{A}$ symbolize a column vector and a matrix, respectively; $\mathbf{A}^T$, $\mathbf{A}^*$, $\mathbf{A}^H$, and $\mathbf{A}^\dag$ stand for the transpose, conjugate, conjugate transpose, and pseudo-inverse of matrix $\mathbf{A}$; The $i$-th row, the $j$-th column, and the $(i,j)$-th entry in matrix $\mathbf{A}$ are denoted as $\mathbf{A}(i,:)$, $\mathbf{A}(:,j)$, and $\mathbf{A}(i,j)$; The determinant, Frobenius norm, and $\ell_p$-norm of matrix $\mathbf{A}$ are expressed as $\det(\mathbf{A})$, $\left\Vert\mathbf{A}\right\Vert_F$, and $||\mathbf{A}||_p$; $\lambda_i(\mathbf{A})$ denotes the $i$-th largest eigenvalue of matrix $\mathbf{A}$, and the corresponding eigenvector is noted as $\boldsymbol{\lambda}_i(\mathbf{A})$; $\tr(\mathbf{A})$ and $\vect(\mathbf{A})$ indicate the trace and vectorization of matrix $\mathbf{A}$; $\circ$ and $\otimes$ stand for the Hadamard and Kronecker products between two matrices; Expectation and the real part of a complex variable are denoted by $\mathbb{E}[\cdot]$ and $\Re[\cdot]$.
	\section{System Model and Problem Formulation}\label{SecII}
	\subsection{System Model}
\begin{figure}[tbp]
		\centering
		\includegraphics[width=16cm]{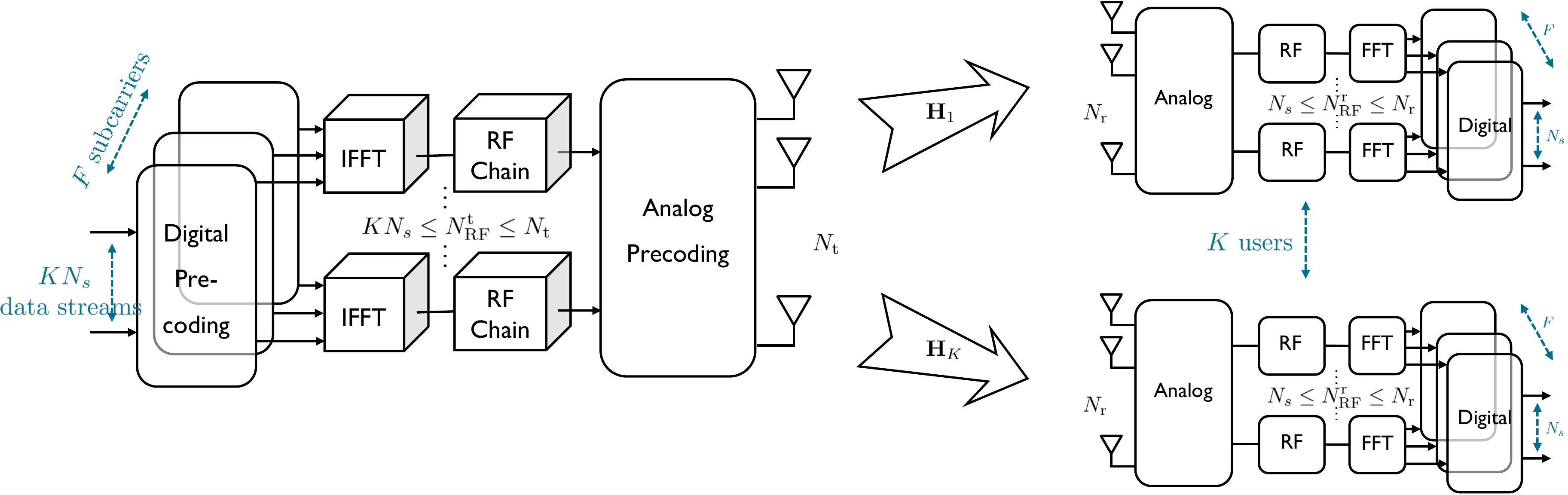}
		\caption{A multiuser OFDM mm-wave MIMO systems with hybrid precoding.}\label{systemmodel}
	\end{figure}

	Consider the downlink transmission of a multiuser OFDM mm-wave MIMO system, as shown in Fig. \ref{systemmodel}, where the base station (BS) is equipped with $N_\mathrm{t}$ antennas and transmits signals to $K$ $N_\mathrm{r}$-antenna users over $F$ subcarriers. On each subcarrier, $N_s$ data streams are transmitted to each user. The limitations of the RF chains are given by $KN_s\le N_\mathrm{RF}^\mathrm{t}< N_\mathrm{t}$ and $N_s\le N_\mathrm{RF}^\mathrm{r}< N_\mathrm{r}$, where $N_\mathrm{RF}^\mathrm{t}$ and $N_\mathrm{RF}^\mathrm{r}$ are the numbers of RF chains facilitated for the BS and each user, respectively.
	
	The received signal for the $k$-th user on the $f$-th subcarrier is given by
	\begin{equation}
	\mathbf{y}_{k,f}={\mathbf{W}^H_\mathrm{BB}}_{k,f}{\mathbf{W}_\mathrm{RF}^H}_k\left(\mathbf{H}_{k,f}\sum_{{k}=1}^K\FRF\FBB_{k,f}\mathbf{s}_{k,f}+\mathbf{n}_{k,f}\right),
	\end{equation}
	where the subscript $k,f$ represents the $k$-th user on the $f$-th subcarrier, and $\mathbf{s}_{k,f}\in\mathbb{C}^{N_s}$ is the transmitted symbol vector such that $\mathbb{E}\left[\mathbf{s}_{k,f}\mathbf{s}^H_{k,f}\right]=\frac{1}{KN_sF}\mathbf{I}_{N_s}$. The digital baseband precoders and combiners are symbolized by $\FBB_{k,f}\in\mathbb{C}^{\NRFt\times N_s}$ and $\WBB_{k,f}\in\mathbb{C}^{\NRFr\times N_s}$, respectively.
	Because the transmitted signals for all the users are mixed together via the digital baseband precoder, and the analog RF precoder is a post-IFFT (inverse fast Fourier transform) operation, the analog RF precoder is shared by all the users and subcarriers, denoted as $\FRF\in\mathbb{C}^{\Nt\times \NRFt}$. Similarly, the analog RF combiner is subcarrier-independent for each user $k$, denoted as $\WRF_k\in\mathbb{C}^{\Nr\times \NRFr}$. Furthermore, the additive noise at the users is represented by $\mathbf{n}_{k,f}\in\mathbb{C}^{\Nr}$, whose elements are independent and identically distributed according to the complex Gaussian distribution $\mathcal{CN}(0, \sigma_n^2)$. The achievable sum rate on the $f$-th subcarrier when transmitted symbols follow a Gaussian distribution is given by {\cite{6717211,5756489}}
	\begin{equation}
	R_{f}=\sum_{k=1}^K\log\det\left(\mathbf{I}_{N_s}+\frac{1}{KN_sF}\mathbf{W}_{k,f}^H\mathbf{H}_{k,f}\mathbf{F}_{k,f}\mathbf{F}_{k,f}^H\mathbf{H}_{k,f}^H\mathbf{W}_{k,f}\mathbf{\Omega}_{k,f}^{-1}\right),
	\end{equation}
	{where
	$\mathbf{F}_{k,f}=\FRF\FBB_{k,f}$ and $\mathbf{W}_{k,f}=\WRF_k\WBB_{k,f}$ are the precoder and combiner matrices, and $\mathbf{\Omega}_{k,f}=\mathbf{W}_{k,f}^H\left[\frac{\rho_k}{KN_sF}\mathbf{H}_{k,f}\left(\sum_{j\ne k}\mathbf{F}_{j,f}\mathbf{F}_{j,f}^H\right)\mathbf{H}_{k,f}^H+\sigma_n^2\right]\mathbf{W}_{k,f}$ stands for the interference plus noise matrix.}
	
	The mm-wave MIMO channel between the BS and the $k$-th user on the $f$-th subcarrier, denoted as $\mathbf{H}_{k,f}$, can be characterized by the Saleh-Valenzuela model as \cite{6717211,7397861,6884253}.
	Although this specific channel model will be used in the simulation, our precoder design approaches are compatible for other general channel models.
	
	\subsection{New Hybrid Precoder Implementation}
	According to the mapping strategies from RF chains to antennas, the hybrid precoder structures can be classified into the fully- and partially-connected ones \cite[Fig. 1]{7397861}. The fully-connected structure fully exploits the degrees of freedom (DoFs) in the RF domain with a natural mapping strategy, i.e., to connect each RF chain to all the antennas. On the contrary, in the partially-connected structure, each antenna element is connected to only one RF chain. These two different mapping strategies (structures) correspond to different constraints in the hybrid precoder design problem, which will be illustrated in detail later in Sections \ref{SecIII} and \ref{SecIV}.
	
\begin{figure}[tbp]
		\centering
		\subfigure[Conventional SPS hybrid precoder implementation.]{
			\includegraphics[height=5cm]{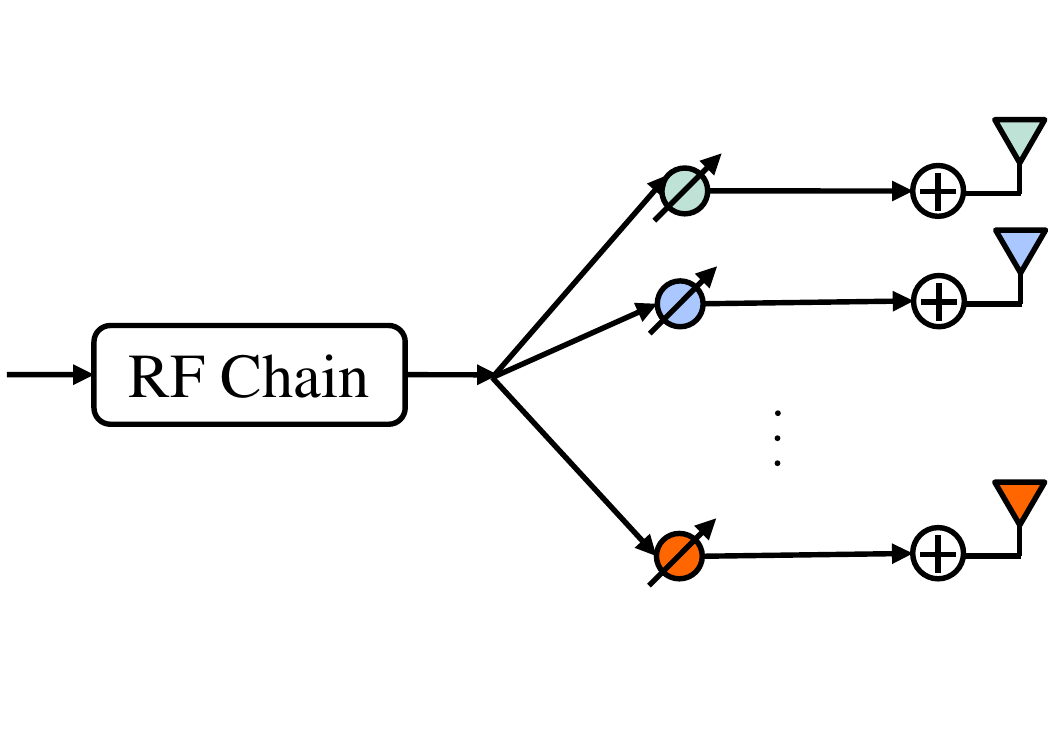}\label{fig11}
		}
		\subfigure[DPS hybrid precoder implementation.]{
			\includegraphics[height=5cm]{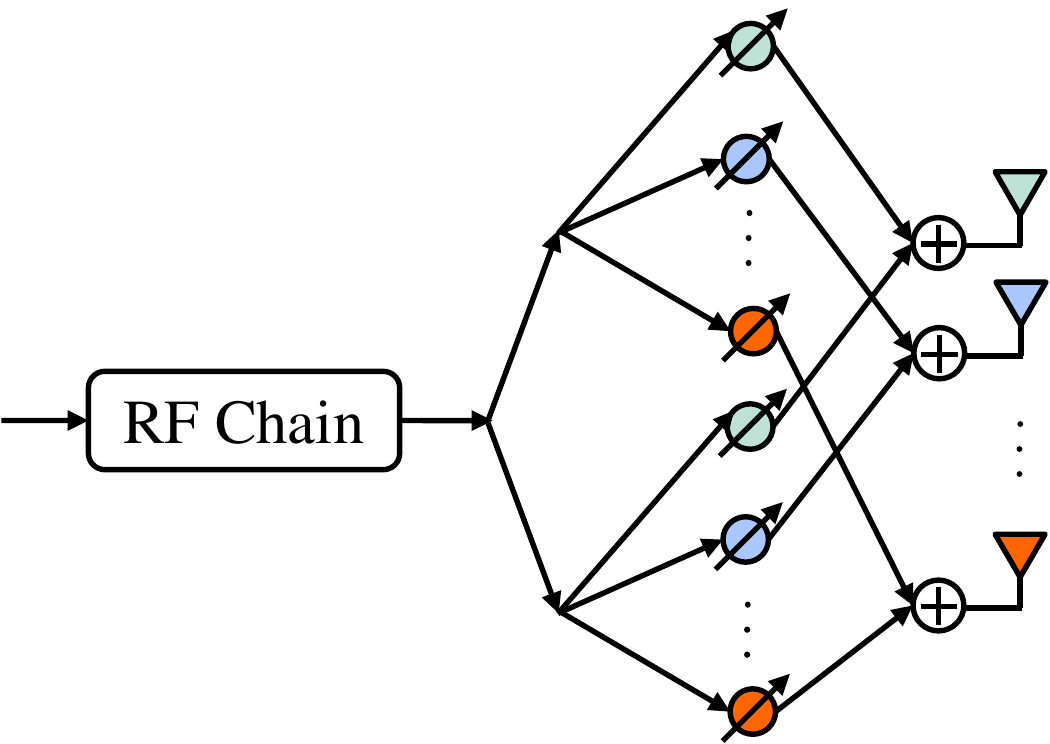}\label{fig12}
		}
		\caption{Comparison of two hybrid precoder implementations. The main difference is on the number of phase shifters in use to compose each connection from an RF chain to a connected antenna element. In the fully-connected structure, each RF chain can be connected to all $\Nt$ antenna elements, and in the partially-connected one, each RF chain is connected to a subset of antennas that do not overlap with each other.}
	\end{figure}

	As mentioned before, the analog RF precoder is practically implemented by phase shifters. Conventionally, in either the fully- or partially-connected structure \cite{7397861}, each connection from a certain RF chain to one of its connected antenna elements is implemented by a single phase shifter, as shown in Fig. \ref{fig11}, which is referred to the \emph{SPS implementation} in this paper. This mapping strategy implies that each non-zero element in the analog precoding and combining matrices should have unit modulus, i.e., $|\FRF(i,j)|=|\WRF(i,j)|=1$. This is intrinsically a non-convex constraint and difficult to tackle, which forms the main design challenge. Although there exist some approaches that can directly deal with this non-convex constraint \cite{7397861,7389996}, the design complexity is still unacceptable in mm-wave systems with much shorter coherent time compared to current sub-6 GHz systems. As a matter of fact, the main obstacle is that we can only adjust the phase but not the amplitude of the RF signals. This motivates us to consider an alternative hybrid precoder implementation which can adjust the amplitude of the RF signals, yet still realized by phase shifters.
	
	In this paper, we propose a new  implementation as shown in Fig. \ref{fig12}, referred as the \emph{DPS implementation} \cite{asilomar}, where the phase shifter network is divided into two groups. For each connection from an RF chain to one of its connected antenna elements, one unique phase shifter in each group will be selected and summed up together to compose the analog precoding gain. With this special implementation, each non-zero element in the analog RF precoding and combining matrices corresponds to a sum of two phase shifters.
	Note that the summation operation creates the possibility to adjust the amplitude of the RF signals, which should be less than two, i.e., the new constraints for the analog RF precoder and combiner are $|\FRF(i,j)|\le 2$ and $|\WRF(i,j)|\le 2$ for all the non-zero entries. By doubling the number of phase shifters, the new constraints become convex and therefore make it more tractable to develop low-complexity design approaches. We impose these amplitude constraints in this paper, and the actual implementation of the phase shifters can then be easily obtained by factorizing a complex number with amplitude less than two into two unit modulus components, expressed as
	\begin{equation}
	ae^{\jmath\theta}=e^{\jmath(\theta+\phi)}+e^{\jmath(\theta-\phi)},
	\end{equation}
	where $a\in[-2,2]$ and $\theta\in[0,2\pi)$ are the amplitude and phase of the non-zero element in $\FRF$ and $\WRF$, and $\phi=\arccos\left(a/2\right)$.
	
{
	\emph{Remark 1:}
	Despite the increased number of phase shifters, as will be shown in this paper, the DPS implementation enjoys unique advantages in both algorithmic and performance aspects. It also provides valuable guidelines for other hybrid precoder design problems. We highlight the benefits of this proposal as follows.
	\begin{itemize}
		\item The DPS implementation greatly simplifies the hybrid precoder design will be greatly simplified when adopting the DPS implementation, , as illustrated in Sections III and IV.
		\item With this new  implementation, hybrid precoders can approach the performance of the fully digital one with fewer RF chains than existing works. Thus, this proposal serves as an algorithmically efficient hybrid precoder design for general multiuser multicarrier mm-wave systems.
		\item The DPS fully-connected hybrid precoder structure serves as a performance upper bound for structures that are with lower hardware complexity. It is a tighter upper bound than the fully digital precoder, especially when the number of RF chains is small.
		\item The precoder design problem becomes a low-rank matrix approximation (eigenvalue) problem for the DPS fully-connected (partially-connected) structure, and theoretical analysis, which is intractable for other structures, becomes possible. It will then help to better understand hybrid precoding systems.
		\item Thanks to the benefits in both performance and algorithmic perspectives, the proposed DPS implementation would drive the hardware research for this implementation.
	\end{itemize}
}	
	
	\subsection{Problem Formulation}
	There exist different  problem formulations for hybrid precoding. Some works tried to directly maximize the spectral efficiency based on approximations and bounds in single-user systems \cite{7389996,7913599}, or based on some extra constraints on the analog precoder to simplify the design in multiuser single-carrier systems \cite{6928432,7335586}. However, when it comes to multiuser multicarrier systems, it is highly challenging and intractable to directly optimize the hybrid precoder with the spectral efficiency being the objective function, given that the spectral efficiency of each user on each subcarrier is coupled with each other by the shared analog RF precoder.
	On the other hand, extensive works showed that minimizing the Euclidean distance\footnote{{In this paper, the Euclidean distance between two precoders refers to the Euclidean distance between two points determined by the vectorization of the two precoding matrices.}} between the fully digital precoder and the hybrid precoder is an effective surrogate for maximizing the spectral efficiency in mm-wave MIMO systems \cite{6717211,7397861,6824962,6874567,6884253,7037444,asilomar}. In this paper, we adopt this alternative objective as our design goal, whose formulation\footnote{In this paper, we focus on the precoder design, and the combiner design problem can be formulated in the same way without the transmit power constraint.} is given by
	\begin{equation}\label{problemformulation}
	\begin{aligned}
	&\underset{\mathbf{F}_\mathrm{RF},\mathbf{F}_\mathrm{BB}}{\mathrm{minimize}} && \left\Vert \mathbf{F}_\mathrm{opt}-\mathbf{F}_\mathrm{RF}\mathbf{F}_\mathrm{BB}\right\Vert _F^2\\
	&\mathrm{subject\thinspace to}&&
	\begin{cases}
	\FRF\in\mathcal{A}\\
	\left\|\mathbf{F}_\mathrm{RF}\mathbf{F}_\mathrm{BB}\right\|_F^2\le KN_sF,
	\end{cases}
	\end{aligned}
	\end{equation}
	where $\Fopt=\left[\Fopt_{1,1},\cdots,\Fopt_{k,f},\cdots,\Fopt_{K,F}\right]\in\mathbb{C}^{\Nt\times KN_sF}$ is the combined fully digital precoder, and $\FBB=\left[\FBB_{1,1},\cdots,\FBB_{k,f},\cdots,\FBB_{K,F}\right]\in\mathbb{C}^{\NRFt\times KN_sF}$ is the concatenated digital baseband precoder. The second constraint is the transmit power constraint at the BS side. The analog RF precoder $\FRF$ is a common component for all $K$ users and $F$ subcarriers, which is restricted in the candidate set $\mathcal{A}\in\{\mathcal{A}_\mathrm{f},\mathcal{A}_\mathrm{d}\}$ induced by the phase shifter implementation. The set $\mathcal{A}$ will be later specified for different hybrid precoder structures.
	{
		Justifications for the formulation \eqref{problemformulation} for single-user systems with flat-fading channels were provided in \cite{6717211}. Here we provide some intuition for this formulation for general hybrid precoding systems. The fully digital precoder serves as a performance upper bound for the hybrid one, and one ideal design goal is to obtain hybrid precoders that approach the performance of the fully digital one. Therefore, it is intuitive to formulate the design problem as approximating the fully digital precoder with the hybrid one.}
	
	With this formulation, the proposed algorithm can be applied with any fully digital precoder. In this paper, we adopt the classical BD precoder as the fully digital one, which is asymptotically optimal in the high signal-to-noise ratio (SNR) regime \cite{1261332}. We will investigate the hybrid precoder design with the DPS implementation for the fully- and partially-connected structures in Sections \ref{SecIII} and \ref{SecIV}, respectively.
	
	\section{Hybrid Precoding for the Fully-connected Structure}\label{SecIII}
	The fully-connected hybrid precoder structure has drawn much research attention in recent years \cite{6717211,7397861,7389996,6874567,6884253,7448873,7335586,7037444,6928432,7913599,zhang2014achieving,7387790}, which will be investigated in this section with the new DPS implementation.
	We will first present an RF-only precoder to demonstrate the advantages of doubling the phase shifters, where the optimization of the analog RF precoder is formulated as a LASSO problem. Afterwards, the hybrid precoder design will be performed via a simple low-rank matrix approximation.
	\subsection{RF-Only Precoding}\label{SecIIIA}
	The main difference between the conventional SPS hybrid precoder implementation and the proposed DPS one is on the analog RF precoder. Therefore, we first present an RF-only precoder design \cite{6851941}, where the analog RF precoder is optimized for a given digital precoder. This problem may arise as a subproblem in hybrid precoder design, as in \cite{asilomar,7037444,7389996}, or for situations where the digital precoder has a fixed design, e.g., from a codebook. The investigation of this problem will demonstrate the algorithmic advantage of the DPS implementation. For the fully-connected structure, the feasible set $\mathcal{A}_\mathrm{f}$ can be specified as {$\mathcal{A}_\mathrm{f}=\left\{\mathbf{A}||\mathbf{A}(i,j)|\le 2\right\}$}, as each RF chain is connected to all the antenna elements. The optimization of the analog RF precoder design problem is given by
	\begin{equation}\label{analogp}
	\begin{aligned}
	&\underset{\mathbf{F}_\mathrm{RF}}{\mathrm{minimize}} && \left\Vert \mathbf{F}_\mathrm{opt}-\mathbf{F}_\mathrm{RF}\mathbf{F}_\mathrm{BB}\right\Vert _F^2\\
	&\mathrm{subject\thinspace to}&&
	\FRF\in\mathcal{A}_\mathrm{f}.
	\end{aligned}
	\end{equation}
	Note that the power constraint in \eqref{problemformulation} is temporarily removed. In fact, after designing the analog RF precoder, we can normalize it if the transmit power constraint is not satisfied. It has been shown in \cite[Lemma 1]{7397861} that as long as we can make the Euclidean distance between the fully digital precoder and the hybrid precoder sufficiently small when ignoring the power constraint, the normalization step will also achieve a small distance to the fully digital precoder.
	The optimization problem \eqref{analogp} is a convex one and can be solved by solvers such as CVX. Nevertheless, to further reduce the computational complexity, we will exploit the inherent structure of the solution by considering its dual problem. 
	\begin{lemma}\label{lem2}
		The dual problem of \eqref{analogp} is a LASSO problem, given by
		\begin{equation}\label{lasso}
		\underset{\mathbf{x}}{\mathrm{minimize}} \quad \frac{1}{2}\left\Vert 
		\mathbf{Ax-b}\right\Vert _2^2+2\Vert\mathbf{x}\Vert_1.
		\end{equation}
		The parameters $\mathbf{A}$ and $\mathbf{b}$ are given by
		\begin{equation}
		\mathbf{A}=\mathbf{S}^{\frac{1}{2}}\mathbf{U},\quad\mathbf{b}=\mathbf{AD}^H\fopt,
		\end{equation}
		where $\mathbf{D}=\mathbf{F}_\mathrm{BB}^T\otimes \mathbf{I}_{\Nt}$ and $\left(\mathbf{D}^H\mathbf{D}\right)^{-1}=\mathbf{USU}^H$ is the singular value decomposition (SVD) of $\left(\mathbf{D}^H\mathbf{D}\right)^{-1}$. The optimal solution of \eqref{analogp} can be written as
		\begin{equation}\label{optimalsolution}
		\mathrm{vec}(\mathbf{F}_\mathrm{RF}^\star)=\mathbf{f}_\mathrm{RF}^\star=\mathbf{A}^H\left(\mathbf{b}-\mathbf{Ax}^\star\right).
		\end{equation}
	\end{lemma}
	\begin{IEEEproof}
		See Appendix \ref{appA}.
	\end{IEEEproof}
	Based on Lemma \ref{lem2}, problem \eqref{analogp} is transferred to a LASSO problem. 
	This provides the opportunity to leverage the large body of existing works on low-complexity algorithms to solve the general LASSO problem \cite{hastie2015statistical}. Recall that, with the conventional SPS implementation, the analog RF precoder is optimized through high-complexity algorithms such as manifold optimization \cite{7397861} to achieve good performance. In contrast, doubling the phase shifters equips us with huge potential to significantly reduce the computational complexity when designing the analog RF precoder.
	
	What deserves an additional mention is a special case where we can get a closed-form solution, which will further reduce the computational complexity.
	It was shown in \cite{7397861,7389996} that a semi-orthogonal structure of the digital baseband precoder, i.e., $\FBB\mathbf{F}_\mathrm{BB}^H=\mathbf{I}_\NRFt$\footnote{Note that in mm-wave multiuser OFDM systems, $\FBB\in\mathbb{C}^{\NRFt\times KN_sF}$, where $KN_sF\ge \NRFt$ for practical system parameters, which means $\FBB$ is a fat matrix.}, leads to an approximately optimal solution. Therefore, we resort to this special case where the observation matrix $\mathbf{A}$ in the LASSO problem \eqref{lasso} is also semi-orthogonal, i.e.,
	\begin{equation}\label{semio}
	\begin{split}
	\mathbf{A}^H\mathbf{A}&=\left(\mathbf{D}^H\mathbf{D}\right)^{-1}=\left((\mathbf{F}_\mathrm{BB}^T\otimes\mathbf{I}_{\NRFt})^H(\mathbf{F}_\mathrm{BB}^T\otimes\mathbf{I}_{\NRFt})\right)^{-1}\\
	&=\left((\mathbf{F}_\mathrm{BB}^*\otimes\mathbf{I}_{\NRFt})(\mathbf{F}_\mathrm{BB}^T\otimes\mathbf{I}_{\NRFt})\right)^{-1}\\
	&=\left((\FBB\mathbf{F}_\mathrm{BB}^H)^T\otimes\mathbf{I}_{\NRFt}\right)^{-1}=\mathbf{I}_{\NRFt^2}.
	\end{split}
	\end{equation}
	With the semi-orthogonal observation matrix $\mathbf{A}$, the LASSO problem \eqref{lasso} has a closed-form solution called soft-thresholding \cite{hastie2015statistical}, which is given by
	\begin{equation}\label{soft}
	\mathbf{x}^\star=\exp\{\jmath\angle(\mathbf{A}^H\mathbf{b})\}\circ\left(\left|\mathbf{A}^H\mathbf{b}\right|-2\right)^+,
	\end{equation}
	where $\angle(\cdot)$, $|\cdot|$, and $(\mathbf{X})^+=\max\{\mathbf{0,X}\}$ are element-wise operations, and the first two extract the phase and amplitude of a complex variable, respectively. Then, substituting \eqref{semio} and \eqref{soft} to \eqref{optimalsolution}, we obtain the corresponding optimal solution to $\FRF$ in \eqref{analogp} as
	\begin{equation}\label{lassoso}
	\mathbf{F}_\mathrm{RF}^\star=\Fopt\mathbf{F}_\mathrm{BB}^H-\exp\left\{\jmath\angle\left(\Fopt\mathbf{F}_\mathrm{BB}^H\right)\right\}\circ\left(\left|\Fopt\mathbf{F}_\mathrm{BB}^H\right|-2\right)^+.
	\end{equation}
	Note that, in order to obtain the optimal analog RF precoder $\mathbf{F}_\mathrm{RF}^\star$ when the digital baseband precoder $\FBB$ is semi-orthogonal, a product of $\Fopt$ and $\mathbf{F}_\mathrm{BB}^H$ is the only required step, which is computationally much more efficient than solving the original problem \eqref{analogp} using an algorithm-embedded solver. This result also suggests that it is beneficial to set the digital baseband precoder as a semi-orthogonal one in the RF-only precoding with the DPS implementation.
	
	\subsection{Hybrid Precoding}
	Previously, we demonstrated the benefit of doubling the phase shifters when optimizing the analog part. When the digital baseband precoder can be jointly optimized, the hybrid precoder design problem is further simplified as an unconstrained matrix decomposition problem, i.e.,
	\begin{equation}\label{formu15}
	\underset{\mathbf{F}_\mathrm{RF},\FBB}{\mathrm{minimize}} \quad\left\Vert \mathbf{F}_\mathrm{opt}-\mathbf{F}_\mathrm{RF}\mathbf{F}_\mathrm{BB}\right\Vert _F^2.\\
	\end{equation}
	
	\emph{Remark 2:}
	The constraint $\FRF\in\mathcal{A}_\mathrm{f}$ in \eqref{analogp}, i.e., $\left|\FRF(i,j)\right|\le2$, is in fact redundant in hybrid precoding. {Once a pair of the unconstrained optimal solution $\left\{\FRF,\FBB\right\}$ is obtained, one can always get another pair of optimal solution $\left\{\frac{\FRF}{\gamma},\gamma\FBB\right\}$ with the factor $\gamma=||\mathrm{vec}(\FRF)||_{\infty}/2$ to satisfy the constraint $\FRF\in\mathcal{A}_\mathrm{f}$, which will not affect the objective value. 
	On the other hand, one may consider deploying $n>2$ phase shifters for each connection from an RF chain to an antenna, and the corresponding constraint would be $|\FRF(i,j)|\le n$. As illustrated above, this constraint is redundant and the factor $\gamma^\prime=||\mathrm{vec}(\FRF)||_\infty/n$ can be applied. Therefore, from both performance and algorithmic perspectives, it does not help to further increase the number of phase shifters. Obviously, the minimum number, i.e., two phase shifters, should be adopted due to cost and power consideration.} 
	Furthermore, the transmit power constraint $\left\|\mathbf{F}_\mathrm{RF}\mathbf{F}_\mathrm{BB}\right\|_F^2\le KN_sF$ is automatically satisfied by the optimal solution of the hybrid precoder, which will be elaborated in the following optimization.
	
	While the main focus of this paper is on multiuser multicarrier systems, some advantages of the proposed DPS implementation in hybrid precoding will be firstly presented in single-carrier systems, as shown in the following result.
	\begin{lemma}\label{lem1}
		For single-carrier systems, with the DPS implementation, the fully digital precoder $\Fopt$ can be perfectly decomposed into  $\FRF$ and $\FBB$ using the minimum number of RF chains, i.e., $\NRFt=KN_s$ and $\NRFr=N_s$. 
	\end{lemma}
	\begin{IEEEproof}
		The proof can be easily obtained by the rank sufficiency of $\FRF$ and $\FBB$ in the decomposition when $F=1$, and is omitted due to space limitation.
	\end{IEEEproof}
	
	Lemma \ref{lem1} shows that, for single-carrier systems with either single-user or multiuser transmissions, the performance of the fully digital precoder can be easily achieved by the hybrid precoder via a simple matrix decomposition. Note that, with the conventional SPS implementation, the number of RF chains should be at least twice that of the data streams in order to realize the fully digital precoder, i.e., $\NRFt=2KN_s$ and $\NRFr=2N_s$ \cite{7397861,7389996}. In this case, since the numbers of phase shifters in use are the same, i.e., $2KN_s\Nt$ at the BS, for both the SPS and DPS implementations, the proposed DPS implementation, which requires fewer RF chains, is more energy efficient when achieving the fully digital precoder.
	
	When it comes to multiuser multicarrier systems, typically $KN_sF\ge \Nt$, the rank of $\Fopt$ should be $\Nt$ (no longer $KN_s$ as single-carrier systems)\footnote{Without loss of generality, we assume all the precoding matrices in \eqref{formu15} have full rank.} and thus perfect decomposition can only be achieved when $\NRFt\ge \Nt$, which, however, severely deviates from the setting of hybrid precoding. Therefore, the matrix decomposition cannot be perfect for hybrid precoder design due to the rank deficiency, i.e., $\NRFt=\mathrm{rank}\left(\FRF\FBB\right)\ll\mathrm{rank}\left(\Fopt\right)=\Nt$. Therefore, problem \eqref{formu15} is typically a low-rank matrix approximation problem, with a closed-form solution as
	\begin{equation}\label{eq16}
	\left(\FRF\FBB\right)^\star\triangleq\mathbf{\hat F}_\mathrm{opt}=\mathbf{U}_1\mathbf{S}_1\mathbf{V}_1^H.
	\end{equation}
	Denote the SVD of $\Fopt$ as $\Fopt=\mathbf{USV}^H$, where matrices $\mathbf{U}_1$ and $\mathbf{V}_1$ are the first $\NRFt$ columns of $\mathbf{U}$ and $\mathbf{V}$, respectively, and $\mathbf{S}_1$ is the diagonal matrix whose diagonal elements are the $\NRFt$ largest singular values of $\Fopt$. This means that the optimal solution of $\FRF\FBB$ is simply obtained by extracting the $\NRFt$ most principle components of $\Fopt$. From the optimal solution \eqref{eq16}, we observe that
	\begin{equation}
	\left\Vert\left(\FRF\FBB\right)^\star\right\Vert_F^2=\left\Vert\mathbf{\hat F}_\mathrm{opt}\right\Vert_F^2\le\left\Vert\Fopt\right\Vert_F^2\le KN_sF,
	\end{equation}
	which means the transmit power constraint is satisfied by the optimal solution $\left(\FRF\FBB\right)^\star$. Until now we have obtained the optimal solution for the entire hybrid precoder, and our next task is to decompose it into two parts. In fact, a large number of options are available for decomposing $\mathbf{\hat F}_\mathrm{opt}$ into $\FRF\FBB$. Nevertheless, we are especially interested in the following one.
	{
	\begin{lemma}\label{lem3}
		The matrix $\mathbf{\hat F}_\mathrm{opt}$ can be decomposed into $\FRF\FBB$ in the following form:
		\begin{equation*}
		\mathbf{F}_\mathrm{RF}=\begin{bmatrix}
		\mathbf{I}_{\NRFt}\\
		\mathbf{\hat F}_\mathrm{opt,2}\mathbf{\hat F}_{\mathrm{opt},1}^\dag
		\end{bmatrix},\quad \mathbf{F}_\mathrm{BB}=\mathbf{\hat F}_{\mathrm{opt},1},
		\end{equation*}
		where $\mathbf{\hat F}_{\mathrm{opt}}=\begin{bmatrix}
		\mathbf{\hat F}_{\mathrm{opt},1}&\mathbf{\hat F}_{\mathrm{opt},2}
		\end{bmatrix}^T$, $\mathbf{\hat F}_{\mathrm{opt},1}$ and $\mathbf{\hat F}_{\mathrm{opt},2}$ are the first $\NRFt$ rows and the $\NRFt+1$-th to $\Nt$-th rows of $\mathbf{\hat F}_\mathrm{opt}$, respectively.
	\end{lemma}
	\begin{IEEEproof}
		Assume $\FRF=\begin{bmatrix}
		\mathbf{I}_\NRFt&\mathbf{X}
		\end{bmatrix}^T$, then the main task to prove Lemma 3 is to find $\mathbf{X}$ and $\FBB$ that satisfy $\FRF\FBB=\mathbf{\hat F}_\mathrm{opt}$.
		
		First, we have
		\begin{equation*}
		\FRF\FBB=\begin{bmatrix}
		\mathbf{I}_\NRFt\\
		\mathbf{X}
		\end{bmatrix}\FBB=\begin{bmatrix}
		\FBB\\
		\mathbf{X}\FBB
		\end{bmatrix}=\mathbf{\hat F}_\mathrm{opt}=\begin{bmatrix}
		\mathbf{\hat F}_{\mathrm{opt},1}\\
		\mathbf{\hat F}_{\mathrm{opt},2}
		\end{bmatrix}.
		\end{equation*}
		Therefore, it is easy to determine that $\FBB=\mathbf{\hat F}_\mathrm{opt,1}$. The remaining task is to solve the equation
		\begin{equation*}
		\mathbf{X}\mathbf{\hat F}_{\mathrm{opt},1}=\mathbf{\hat F}_\mathrm{opt,2}.
		\end{equation*}
		Since $\mathbf{\hat F}_\mathrm{opt}$ is with rank $\NRFt$ and is obtained by the SVD of $\Fopt$, the first $\NRFt$ rows of $\mathbf{\hat F}_\mathrm{opt}$ (the rows of $\mathbf{\hat F}_{\mathrm{opt},1}$) are linearly independent, and the remaining rows in $\mathbf{\hat F}_\mathrm{opt}$ (the rows of $\mathbf{\hat F}_{\mathrm{opt},2}$) can be linearly expressed by the rows of $\mathbf{\hat F}_{\mathrm{opt},1}$. Hence, $\mathbf{X}=\mathbf{\hat F}_\mathrm{opt,2}\mathbf{\hat F}_{\mathrm{opt},1}^\dag$ is the solution to the equation, which completes the proof.
	\end{IEEEproof}
}
	The advantage of this decomposition form lies in the pattern of $\FRF$ in Lemma \ref{lem3}. The first $\NRFt$ rows of $\FRF$ form an identity matrix, which in fact does not need a phase shifter implementation since the zero elements correspond to no connections whereas the diagonal elements refer to direct connections from RF chains to antennas. This means we only need $2\NRFt(\Nt-\NRFt)$ phase shifters in the analog RF precoder, instead of $2\NRFt\Nt$. Although a similar result was presented in \cite{7387790}, note that the result in Lemma \ref{lem3} saves $2\NRFt$ more phase shifters and the method is simpler and more straightforward than the decomposition procedure involving two QR decompositions as in \cite{7387790}. 
	Furthermore, the decomposition pattern in Lemma \ref{lem3} can also be applied to single-carrier systems based on the result in Lemma \ref{lem1}, which will further improve the energy efficiency when achieving the fully digital precoder.
	
	As demonstrated above, by doubling the phase shifters, what we need for the hybrid precoder design is computing a subset of singular values and vectors of the fully digital precoding matrix, i.e., the $\NRFt$ most principle components of $\Fopt$, whose computational complexity is $\mathcal{O}\left(KN_sF\Nt\NRFt\right)$. Recall that OMP, as the most popular algorithm for the conventional SPS implementation, is with the computational complexity $\mathcal{O}\left(KN_sF\Nt\NRFt\left(KL+\NRFt^2\right)\right)$, which is higher than that of the simple approach we proposed and is related to the channel parameter $L=\sum_{k=1}^KN_{\mathrm{cl},k}N_{\mathrm{ray},k}$. In other words, the proposed DPS implementation equips us with precoding algorithms computationally much more efficient than existing ones. Later in Section \ref{SecV}, its merits on achievable performance will also be demonstrated via simulations.
	
	{
	\subsection{DPS-Enabled SPS Hybrid Precoding}\label{SecIIID}
	In this part, inspired by the above hybrid precoder design, we propose an efficient way to design the conventional SPS implementation. In particular, based on the solution for the DPS implementation, we adopt an heuristic way to tackle the unit modulus constraints induced by the SPS implementation. 
	
	 As shown in \eqref{eq16}, the optimal hybrid precoder $\mathbf{\hat{F}_\mathrm{opt}}$ can be decomposed by SVD. Therefore, one optimal solution to the hybrid precoder with the DPS implementation is
	\begin{equation}
	\FRF=\mathbf{U}_1,\quad\FBB=\mathbf{S}_1\mathbf{V}_1^H.
	\end{equation}
	 Note that the unitary matrix $\mathbf{U}_1$ fully extracts the information of the column space of $\mathbf{\hat{F}_\mathrm{opt}}$, whose basis are the orthonormal columns in $\FRF$.
	 
	 In contrast, in the SPS implementation, the unit modulus constraints, i.e., $\left|\FRF\left(i,j\right)\right|=1$ not only require each column in $\FRF$ to have a constant norm like $\mathbf{U}_1$, but also induce an element-wise constraint. Since each element in $\FRF$ can only contain the phase information, we propose to extract the phases of the optimal analog precoder for the DPS implementation to construct the SPS solution, given by
	 \begin{equation}
	 \FRF=\exp\{\jmath\angle\left(\mathbf{U}_1\right)\}.
	 \end{equation}
	Although this step is based on heuristics, it shall be shown in Section \ref{SecV} that simply extracting the phase information only incurs negligible performance loss. Similar approaches can also be found in \cite{6928432,7397861}.
	
	Compared with existing hybrid precoding algorithms with the SPS implementation, e.g., the MO-AltMin \cite{7397861}, OMP \cite{6717211} algorithms, and the algorithm in \cite{7913599}, the proposed DPS-enabled design method enjoys much lower computational complexity without any iterative procedure, which makes it a good candidate for low-complexity hybrid precoding with the SPS fully-connected structure. 	
}
	
	\subsection{Interuser Interference Cancellation}\label{SecIIIC}
	While we can perfectly cancel the interuser interference with the fully digital precoder $\Fopt$, there will be residual interuser interference when applying the hybrid precoder, which is an approximation of the fully digital one. For the same reason, as hybrid combining is adopted at the receiver side, the interuser interference cannot be canceled by the receiver either. Later in Section \ref{SecV}, we will see that in multiuser multicarrier systems, interuser interference is a severe problem that will dramatically degrade the hybrid precoding performance, especially at high SNRs.
	
	In this subsection, after designing the hybrid precoder and combiner, we propose to cascade another digital baseband precoder $\FBD$ that is responsible for canceling the residual interuser interference. In particular, with the hybrid precoder and combiner at hand, we define an effective channel for the $k$-th user on the $f$-th subcarrier as
	\begin{equation}\label{effch}
	\mathbf{\hat H}_{k,f}={\mathbf{W}^H_\mathrm{BB}}_{k,f}{\mathbf{W}^H_\mathrm{RF}}_{k}\mathbf{H}_{k,f}{\FRF}{\FBB}_{f},
	\end{equation}
	where $\FBB_f=\left[\FBB_{1,f},\cdots,\FBB_{k,f},\cdots,\FBB_{K,f}\right]\in\mathbb{C}^{\NRFt\times KN_s}$ is the composite digital precoder on the $f$-th subcarrier. Our goal is to design the precoders $\FBD_{k,f}$, which satisfy the conditions
	\begin{equation}\label{eq21}
	\mathbf{\hat H}_{j,f}\FBD_{k,f}=\mathbf{0}, \quad k\ne j.
	\end{equation}
	A simple way to achieve the conditions is the BD precoder, and note that the dimension of the effective channel is $N_s\times KN_s$, which is sufficient for BD. More details can be found in \cite{1261332}. Therefore, after cascading the BD precoder at the baseband, the overall digital baseband precoder of the $k$-th user on the $f$-th subcarrier is
	\begin{equation}
	{{\mathbf{F}}_\mathrm{B}}_{k,f}=\FBB_{k,f}\FBD_{k,f}.
	\end{equation}
	Since now we have obtained an interuser interference free system, we can normalize the precoder to satisfy the maximum transmit power, in order to improve the SNRs of the users. The same approach to cancel the interuser interference will also be used in the partially-connected structure and will not be repeatedly presented in the next section.

	\section{Hybrid Precoding for the Partially-connected Structure}\label{SecIV}
	One of the shortages of the fully-connected structure is the large number of phase shifters. The partially-connected structure, as a more energy efficient and cost-effective structure \cite{7397861,7445130}, employs notably fewer phase shifters, i.e., $2\Nt$ phase shifters with the DPS implementation, which lends itself to practical implementation. Since the DoFs of the analog precoder is greatly reduced, RF-only precoding is far from satisfactory in the partially-connected structure. In this section, we shall first present the hybrid precoding with a fixed mapping from RF chains to antennas. Two algorithms will be then proposed to perform dynamic mapping to further improve the  performance.
	
	\subsection{Hybrid Precoding With Fixed Mapping}\label{IVA}
	In \cite{7397861,7445130}, fixed mapping was considered in the partially-connected structure, i.e., each RF chain is connected to a certain number of antennas in a predetermined manner.
	To present the hybrid precoder design with fixed mapping clearly, we take one special mapping \cite{7397861,7445130,7370753} as an example in the following, where the $j$-th RF chain is connected to the $j$-th set of $\Nt/\NRFt$ adjacent antennas. The corresponding constraint on the analog RF precoding matrix can be visualized as a set of block diagonal matrices $\mathcal{A}_\mathrm{b}$, where each block is an $\Nt/\NRFt$ dimension vector, i.e.,
	\begin{equation}\label{eq22}
	\FRF=\begin{bmatrix}
	\mathbf{p}_1&\mathbf{0}&\cdots&\mathbf{0}\\
	\mathbf{0}&\mathbf{p}_2&&\mathbf{0}\\
	\vdots&&\ddots&\vdots\\
	\mathbf{0}&\mathbf{0}&\cdots&\mathbf{p}_\NRFt
	\end{bmatrix},
	\end{equation}
	where $\mathbf{p}_j=\left[a_{(j-1)\frac{\Nt}{\NRFt}+1},\cdots,a_{j\frac{\Nt}{\NRFt}}\right]^T$. The amplitude of the analog precoding gain for the $i$-th connection from RF chains to antennas is denoted as $|a_i|\le2$. Similar to the hybrid precoding in the DPS fully-connected structure, the constraints $|a_i|\le2$ are redundant and therefore they are omitted in the following derivation. Furthermore, the transmit power constraint $\left\|\mathbf{F}_\mathrm{RF}\mathbf{F}_\mathrm{BB}\right\|_F^2\le KN_sF$ is also automatically satisfied by the optimal solution of the hybrid precoder, which will be shown in the following parts. Thus, the hybrid precoder design problem with fixed mapping can be recast as
	\begin{equation}\label{eq23}
	\begin{aligned}
	&\underset{\mathbf{F}_\mathrm{RF},\FBB}{\mathrm{minimize}} && \left\Vert \mathbf{F}_\mathrm{opt}-\mathbf{F}_\mathrm{RF}\mathbf{F}_\mathrm{BB}\right\Vert _F^2\\
	&\mathrm{subject\thinspace to}&&
	\FRF\in\mathcal{A}_\mathrm{b}.
	\end{aligned}
	\end{equation} 
	Note that there is only one non-zero element in each row of the analog RF precoding matrix $\FRF$. Due to this special structure, different vectors $\mathbf{p}_j$ will be multiplied by distinct rows of $\FBB$, which decouples problem \eqref{eq23} into $\NRFt$ subproblems in an RF chain-by-RF chain sense. The optimization of the hybrid precoder for the $j$-th RF chain is given by
	\begin{equation}
	\mathcal{P}_j:\quad\underset{\{a_i\},\mathbf{x}_j}{\mathrm{minimize}}  \sum_{i\in\mathcal{F}_j}
	\left\Vert\mathbf{y}_i-a_i\mathbf{x}_j\right\Vert_2^2,\\
	\end{equation}
	where $\mathcal{F}_j=\left\{i\in\mathbb{Z}\left|{(j-1)\frac{\Nt}{\NRFt}+1\le i\le j\frac{\Nt}{\NRFt}}\right.\right\}$, $\mathbf{y}_i=\mathbf{F}_\mathrm{opt}^T(i,:)$, and $\mathbf{x}_j=\mathbf{F}_\mathrm{BB}^T(j,:)$. 
	\begin{prop}
		The optimal solution to the subproblem $\mathcal{P}_j$ is given by the following closed-form expression.
		\begin{equation}
		\mathbf{x}_j^\star=\boldsymbol{\lambda}_1\left(\sum_{i\in\mathcal{F}_j}\mathbf{y}_i\mathbf{y}_i^H\right),\quad a_i^\star=\frac{\mathbf{x}_j^H\mathbf{y}_i}{||\mathbf{x}_j||_2^2}.
		\end{equation}
		\begin{IEEEproof}
			We check the first order optimality conditions as
			\begin{align}
			&\frac{\partial}{\partial a_i}f(a_i,\mathbf{x}_j)=0{\Rightarrow-\mathbf{y}_i^H\mathbf{x}_j+a_i^*||\mathbf{x}_j||^2_2 =0}\Rightarrow a_i=\frac{\mathbf{x}_j^H\mathbf{y}_i}{||\mathbf{x}_j||_2^2}\label{eq3}\\
			&\frac{\partial}{\partial \mathbf{x}_j}f(a_i,\mathbf{x}_j)=\mathbf{0}{\Rightarrow\sum_{i\in\mathcal{F}_j}-a_i\mathbf{y}_i^H+|a_i|^2\mathbf{x}_j^H=\mathbf{0}}\Rightarrow \sum_{i\in\mathcal{F}_j}|a_i|^2\mathbf{x}_j=\sum_{i\in\mathcal{F}_j}a_i^*\mathbf{y}_i.,\label{eq4}
			\end{align}
			where $f(a_i,\mathbf{x}_j)$ is the objective function in subproblem $\mathcal{P}_j$.
			Substituting \eqref{eq3} into \eqref{eq4}, we can get 
			\begin{equation}
			\begin{split}
			&\mathbf{x}_j^H\sum_{i\in\mathcal{F}_j}|a_i|^2=\frac{\mathbf{x}_j^H}{||\mathbf{x}_j||_2^2}\sum_{i\in\mathcal{F}_j}\mathbf{y}_i\mathbf{y}_i^H\\
			\Rightarrow&\left(\sum_{i\in\mathcal{F}_j}\mathbf{y}_i\mathbf{y}_i^H\right)\mathbf{x}_j=\left(||\mathbf{x}_j||_2^2\sum_{i\in\mathcal{F}_j}|a_i|^2\right)\mathbf{x}_j
			\triangleq\lambda_j\mathbf{x}_j,
			\end{split}
			\end{equation}
			which shows that $\lambda_j$ and $\mathbf{x}_j$ are the eigenvalue and eigenvector of $\sum_{i\in\mathcal{F}_j}\mathbf{y}_i\mathbf{y}_i^H$.
			Moreover, by substituting \eqref{eq3} into the objective function in $\mathcal{P}_j$, it can be rewritten as
			\begin{equation}\label{eq28}
			f(a_i,\mathbf{x}_j)=\sum_{i\in\mathcal{F}_j}\mathbf{y}_i^H\mathbf{y}_i-|a_i|^2\mathbf{x}_j^H\mathbf{x}_j=\sum_{i\in\mathcal{F}_j}||\mathbf{y}_i||_2^2-\lambda_j.
			\end{equation}
			Hence, minimizing the objective function is equivalent to taking $\lambda_j$ as the largest eigenvalue of the covariance matrix $\sum_{i\in\mathcal{F}_j}\mathbf{y}_i\mathbf{y}_i^H$, denoted as $\lambda_1\left(\sum_{i\in\mathcal{F}_j}\mathbf{y}_i\mathbf{y}_i^H\right)$. 
		\end{IEEEproof}
	\end{prop}

	From equation \eqref{eq28}, we  obtain 
	\begin{equation}\label{eq29}
	\begin{split}
	\left\Vert\Fopt-\FRF\FBB\right\Vert_F^2&=\left\Vert\Fopt\right\Vert_F^2-\sum_{j=1}^\NRFt\lambda_j
	=\left\Vert\Fopt\right\Vert_F^2-\sum_{j=1}^\NRFt\left(||\mathbf{x}_j||_2^2\sum_{i\in\mathcal{F}_j}|a_i|^2\right)\\
	&=\left\Vert\Fopt\right\Vert_F^2-\left\Vert\FRF\FBB\right\Vert_F^2\ge0\Rightarrow\left\Vert\FRF\FBB\right\Vert_F^2\le\left\Vert\Fopt\right\Vert_F^2\le KN_sF,
	\end{split}
	\end{equation}
	which means that the transmit power constraint is naturally satisfied by the optimal solutions.
	While we fixed the mapping strategy as shown in \eqref{eq22}, the proposed design approach is applicable to an arbitrary mapping strategy.
	
	\subsection{Hybrid Precoding With Dynamic Mapping}
	Different from the fully-connected structure that utilizes all the connections from RF chains to antennas, 
	the partially-connected structure will induce non-negligible performance loss \cite{7397861}. In this section, we propose to improve its performance  by optimizing the mapping strategy, i.e., we will dynamically determine for each RF chain which antennas it should be connected. The dynamic mapping problem is given as
	\begin{equation}
	\begin{aligned}
	&\underset{\mathbf{F}_\mathrm{RF},\FBB}{\mathrm{minimize}} && \left\Vert \mathbf{F}_\mathrm{opt}-\mathbf{F}_\mathrm{RF}\mathbf{F}_\mathrm{BB}\right\Vert _F^2\\
	&\mathrm{subject\thinspace to}&&
\FRF\in\mathcal{A}_\mathrm{d},
	\end{aligned}
	\end{equation} 
	where $\mathcal{A}_\mathrm{d}$ is a set of matrices for which every row only has one non-zero entry, i.e., {$\mathcal{A}_\mathrm{d}=\left\{\mathbf{A}|||\mathbf{A}(i,:)||_0=1\right\}$}, meaning that each antenna can only be connected to one RF chain.
	As indicated by equation \eqref{eq29}, once the mapping is fixed, the optimal value of the objective function in \eqref{eq23} is 
	\begin{equation}
	\left\Vert\Fopt\right\Vert_F^2-\sum_{j=1}^\NRFt\lambda_1\left(\sum_{i\in\mathcal{D}_j}\mathbf{y}_i\mathbf{y}_i^H\right).
	\end{equation}
	Hence, when we have the freedom to design the mapping strategy from RF chains to antennas, the design target is to seek the mapping that maximizes the sum of the largest eigenvalues, i.e.,
	\begin{equation}\label{eq31}
	\begin{aligned}
	&\underset{\{\mathcal{D}_j\}_{j=1}^\NRFt}{\mathrm{maximize}} && \sum_{j=1}^\NRFt\lambda_1\left(\sum_{i\in\mathcal{D}_j}\mathbf{y}_i\mathbf{y}_i^H\right)\\
	&\mathrm{subject\thinspace to}&&
	\begin{cases}
	\cup_{j=1}^\NRFt\mathcal{D}_j=\left\{1,\cdots,\Nt\right\}\\
	\mathcal{D}_j\cap\mathcal{D}_k=\emptyset,\quad\forall j\ne k,
	\end{cases}
	\end{aligned}
	\end{equation} 
	where $\mathcal{D}_j$ is the mapping set containing the antenna indices that are mapped to the $j$-th RF chain. The dynamic mapping problem is a combinatorial problem and the optimal solution can be given by exhaustive search with an extremely huge number of possible mapping strategies as $\frac{1}{(\NRFt)!}\sum_{k=0}^\NRFt(-1)^{\NRFt-k}\binom{\NRFt}{k}k^\Nt$, which prevents its practical implementation. Therefore, we first propose a greedy algorithm to solve the problem. {The pseudocode of the algorithm is omitted due to its simplicity and space limitation.}
	
	In each iteration of the greedy algorithm, we connect the ${p^\star}$-th antenna to the $j^\star$-th RF chain, which is the connection with the maximum increment of the largest eigenvalue when this connection is added into the mapping network. Note that the computational complexity of the algorithm is dominated by the calculation of the largest eigenvalue. In the greedy algorithm, the number of times we need to perform the eigenvalue decomposition (EVD) is $\mathcal{O}\left(\Nt\NRFt(1+\Nt)/2\right)$, which is a quite large number especially when large-scale antenna arrays are leveraged in mm-wave MIMO systems. To relieve us from the high computational complexity, we then propose a modified K-means algorithm to solve the dynamic design problem \eqref{eq31}.
	
	We reconsider problem \eqref{eq31} as follows. The problem is equivalent to classifying $\Nt$ vectors (antennas) into $\NRFt$ clusters (RF chains). K-means, aiming at partitioning the observation vectors into $K_\mathrm{cl}$ clusters, is a prevalent approach for cluster analysis in data mining, where $K_\mathrm{cl}$ is a predefined parameter, and turns out to be suitable for problem \eqref{eq31}. In the classical K-means algorithm, the objective is to minimize the sum of the Euclidean distances from each observation vector to the centroid of the cluster it belongs to. The distortion function that is to be minimized in the classical K-means algorithm is given by\footnote{We present the distortion function in the classified K-means algorithm with a slight abuse of notations $\mathbf{y}_i$ and $\mathbf{x}_j$ so that the content of the modified one in the following is easier to follow.}
	\begin{equation}
	D(\mathbf{y}_i,\mathbf{x}_j)=\sum_{j=1}^{K_\mathrm{cl}}\sum_{i\in\mathcal{K}_j}||\mathbf{y}_i-\mathbf{x}_j||_2^2,
	\end{equation}
	where $\mathbf{y}_i$ are the observation vectors while $\mathbf{x}_j$ is the centroid of the $j$-th cluster.
	
	However, this distortion function cannot be directly adopted to solve the dynamic mapping design problem \eqref{eq31} since the objectives are quite different. In \eqref{eq31}, the objective is to maximize the sum of the largest eigenvalues of the covariance matrices of each cluster. Therefore, we propose to modify the distortion function in the K-means algorithm as
	\begin{equation}\label{eq33}
	D^\prime(\mathbf{y}_i,\mathbf{x}_j)=\sum_{j=1}^\NRFt\frac{\mathbf{x}_j^H\left(\sum_{i\in\mathcal{D}_j}\mathbf{y}_i\mathbf{y}_i^H\right)\mathbf{x}_j}{\mathbf{x}_j\mathbf{x}_j^H}.
	\end{equation}
	The modified distortion function is the sum of Rayleigh quotients of the covariance matrices of each cluster, whose optimal value is the sum of the largest eigenvalues when we maximize \eqref{eq33} over $\mathbf{x}_j$. The overall clustering problem can be written as
	\begin{equation}\label{eq34}
	\begin{aligned}
	&\underset{\{\mathcal{D}_j,\mathbf{x}_j\}_{j=1}^\NRFt}{\mathrm{maximize}} && \sum_{j=1}^\NRFt\frac{\mathbf{x}_j^H\left(\sum_{i\in\mathcal{D}_j}\mathbf{y}_i\mathbf{y}_i^H\right)\mathbf{x}_j}{\mathbf{x}_j\mathbf{x}_j^H}\\
	&\mathrm{subject\thinspace to}&&
	\begin{cases}
	\cup_{j=1}^\NRFt\mathcal{D}_j=\left\{1,\cdots,\Nt\right\}\\
	\mathcal{D}_j\cap\mathcal{D}_k=\emptyset,\quad\forall j\ne k.
	\end{cases}
	\end{aligned}
	\end{equation} 
	We propose to adopt alternating maximization to solve this problem, which alternately updates the clustering and centroids when the other one is fixed. This approach results in closed-form solutions for the two update procedures.
	
	In the clustering update, we allocate each vector to the cluster whose centroid has the largest inner product with it, i.e., allocate $\mathbf{y}_i$ to the $j^\star$-th cluster, where
	\begin{equation}\label{clustering}
	j^\star =\arg\underset{j}{\max}\quad\left|\mathbf{y}_i^H\mathbf{x}_j\right|^2.
	\end{equation}
	In the centroid update, the optimization of the centroids is equivalent to maximizing  the Rayleigh quotients for each cluster, whose optimal solution is simply given by the eigenvector corresponding to the largest eigenvalue, i.e.,
	\begin{equation}\label{centroid}
	\mathbf{x}_j^\star=\boldsymbol{\lambda}_1\left(\sum_{i\in\mathcal{D}_j}\mathbf{y}_i\mathbf{y}_i^H\right).
	\end{equation}
	Now we have the modified K-means algorithm, which is summarized as \textbf{Algorithm 1}.
	\begin{algorithm}[t]
		\caption{Modified K-means Algorithm for Dynamic Mapping in the Partially-connected Structure}
		\begin{algorithmic}[1]
			\REQUIRE
			$\left\{\mathbf{y}_i=\mathbf{F}_\mathrm{opt}^T(i,:)\right\}_{i=1}^\Nt$
			\STATE Construct the initial centroids $\{\mathbf{x}_j\}_{j=1}^\NRFt$;
			\REPEAT
			\STATE Fix the centroids, allocate $\{\mathbf{y}_i\}_{i=1}^\Nt$ to the clusters according to \eqref{clustering};
			\STATE Optimize the centroids $\{\mathbf{x}_j\}_{j=1}^\NRFt$ using \eqref{centroid} when the clustering is fixed;
			\UNTIL convergence;
			\STATE Calculate $\mathbf{F}_\mathrm{BB}^T(j,:)=\mathbf{x}_j=\boldsymbol{\lambda}_1\left(\sum_{i\in\mathcal{D}_j}\mathbf{y}_i\mathbf{y}_i^H\right)$ for $j\in\{1,\cdots,\NRFt\}$;
			\STATE Compute the effective channels $\mathbf{\hat H}_{k,f}={\mathbf{W}^H_\mathrm{BB}}_{k,f}$ and BD precoders $\FBD_{k,f}$ according to \eqref{effch} and \eqref{eq21};
			\STATE  For the digital precoder at the transmit end, normalize
			$\widehat{\mathbf{F}}_\mathrm{B}=\frac{\sqrt{KN_sF}}{\left\Vert\mathbf{F}_\mathrm{RF}\mathbf{F}_\mathrm{B}\right\Vert_F}\mathbf{F}_\mathrm{B}$.
		\end{algorithmic}
	\end{algorithm}
	
	Note that Steps 3 and 4 both give the globally optimal solutions to the clustering and centroid. Hence, the algorithm will converge to a stationary point since it is a two block coordinate descent procedure \cite{grippo2000convergence}. Because the modified distortion function is not jointly convex with respect to $\left\{\mathcal{D}_j\right\}_{j=1}^\NRFt$ and $\left\{\mathbf{x}_j\right\}_{j=1}^\NRFt$, the modified K-means algorithm converges to a local optimum of problem \eqref{eq34} so the solution is sensitive to the initial centroids selection. 
	For hybrid precoding, the size of the observation set is much larger than the cluster number, i.e., $\Nt\gg\NRFt$. One heuristic rule of thumb to design the initial centroids is to pick $\NRFt$ observation vectors with small correlations. 
	In our proposed algorithm, we propose to select $\NRFt/2$ pairs of vectors out of the $\Nt$ observation vectors as the initial centroids, which have the $\NRFt/2$ smallest inner products. 
	
	Recall that EVD is the dominant part of the computational complexity in dynamic mapping design. In each alternating iteration in the modified K-means algorithm, $\NRFt$ times of EVD are needed and therefore the overall times are $\mathcal{O}(N\NRFt)$, where $N$ is the iteration number. For practical settings in Section \ref{SecV}, the modified K-means algorithm typically converges within 10 iterations, which is much less than $\Nt(1+\Nt)/2$ and thus results in significant complexity reduction compared to the greedy algorithm.
	
	\subsection{Fully-connected vs. Partially-connected Structures}\label{SecIVC}
	There exist several studies \cite{7335586,7397861,7389996} investigating different design algorithms for the fully- and partially-connected structures, and comparisons between these two structures are provided via simulations. 
	However, to the best of the authors' knowledge, so far there is no analytical quantitative comparisons between different structures. The complicated design approaches to handle the unit modulus constraints induced by the SPS implementation are the main obstacles. With the DPS implementation and its resulting low-complexity design approaches at hand, we are able to fill this gap. Following \eqref{eq16} and \eqref{eq23} in Sections \ref{SecIII} and \ref{SecIV}, we obtain that
	\begin{equation}
	f^\star_\mathrm{f}=\sum_{p=\NRFt+1}^\Nt\sigma_p^2\left(\Fopt\right),\quad
	f^\star_\mathrm{p}=\left\Vert\Fopt\right\Vert_F^2-\sum_{j=1}^\NRFt\lambda_1\left(\sum_{i\in\mathcal{D}_j}\mathbf{y}_i\mathbf{y}_i^H\right),
	\end{equation}
	where $f^\star_\mathrm{f}$ and $f^\star_\mathrm{p}$ are the optimal values of the objective function in \eqref{problemformulation} for the fully- and partially-connected structures, respectively, and $\sigma_p(\Fopt)$ denotes the $p$-th largest singular value of $\Fopt$. We define the performance gap $\Delta$ between the fully- and partially-connected structures as the different between two objective values, i.e.,
	\begin{equation}\label{distance}
	\begin{split}
	\Delta&\triangleq f^\star_\mathrm{p}-f^\star_\mathrm{f}=\left\Vert\Fopt\right\Vert_F^2-\sum_{j=1}^\NRFt\lambda_1\left(\sum_{i\in\mathcal{D}_j}\mathbf{y}_i\mathbf{y}_i^H\right)-\sum_{p=\NRFt+1}^\Nt\sigma_p^2\left(\Fopt\right)\\
	&=\sum_{p=1}^\NRFt\lambda_p\left(\mathbf{F}_\mathrm{opt}^H\Fopt\right)-\sum_{j=1}^\NRFt\lambda_1\left(\sum_{i\in\mathcal{D}_j}\mathbf{y}_i\mathbf{y}_i^H\right)=\sum_{p=1}^\NRFt\lambda_p\left(\mathbf{F}_\mathrm{opt}^H\Fopt\right)-\sum_{j=1}^\NRFt\lambda_1\left(\mathbf{Y}_j\mathbf{Y}_j^H\right),
	\end{split}
	\end{equation} 
	where $\mathbf{Y}_j$ is composed of the vectors $\left\{\mathbf{y}_i\right\}_{i\in\mathcal{D}_j}$ as its columns.   Therefore, once the RF-antenna mapping in the partially-connected structure is determined, the performance gap is given by \eqref{distance}, which provides an analytical comparison of two hybrid precoder structures.
	This expression indicates that the performance gap depends on the channel realization, as well as the RF chain-antenna mapping strategy. 
	
	\section{Simulation Results}\label{SecV}
	In this section, we numerically evaluate the performance of the proposed hybrid precoder design for multiuser OFDM mm-wave MIMO systems, with $F=128$ subcarriers are assumed. 
	The channel parameters are given by $N_{\mathrm{cl},k}=3$ clusters, $N_{\mathrm{ray},k}=8$ rays, and the average power of each cluster is $\sigma_\alpha^2=1$. The AoDs and AoAs follow the Laplacian distribution with uniformly distributed mean angles in $[0,2\pi)$ and angular spread of 10 degrees.
	The antenna elements in the USPA are separated by half wavelength, and all simulation results are averaged over 1000 channel realizations.
	
	\subsection{RF-Only Precoding in the Fully-Connected Structure}
	First, we test how much performance gain we can get when we double the number of phase shifters via investigating the RF-only precoding in the fully-connected structure. Note that, with the conventional SPS implementation, manifold optimization was shown in \cite{7397861} to be an effective method to directly tackle the unit modulus constraints and achieve higher spectral efficiency than other existing works. In this subsection, we adopt Algorithm 1 in \cite{7397861} for the SPS implementation as the benchmark to show the advantage of the proposed DPS implementation. For fair comparison, we always keep the same digital baseband precoder $\FBB$ that is semi-orthogonal for both SPS and DPS implementations in the simulation\footnote{Note that we cannot adjust the digital precoder in the RF-only precoding so we do not apply the additional BD mentioned in Section \ref{SecIIIC} at this stage.}.
	\begin{figure}[tbp]
		\centering
		\includegraphics[height=5.75cm]{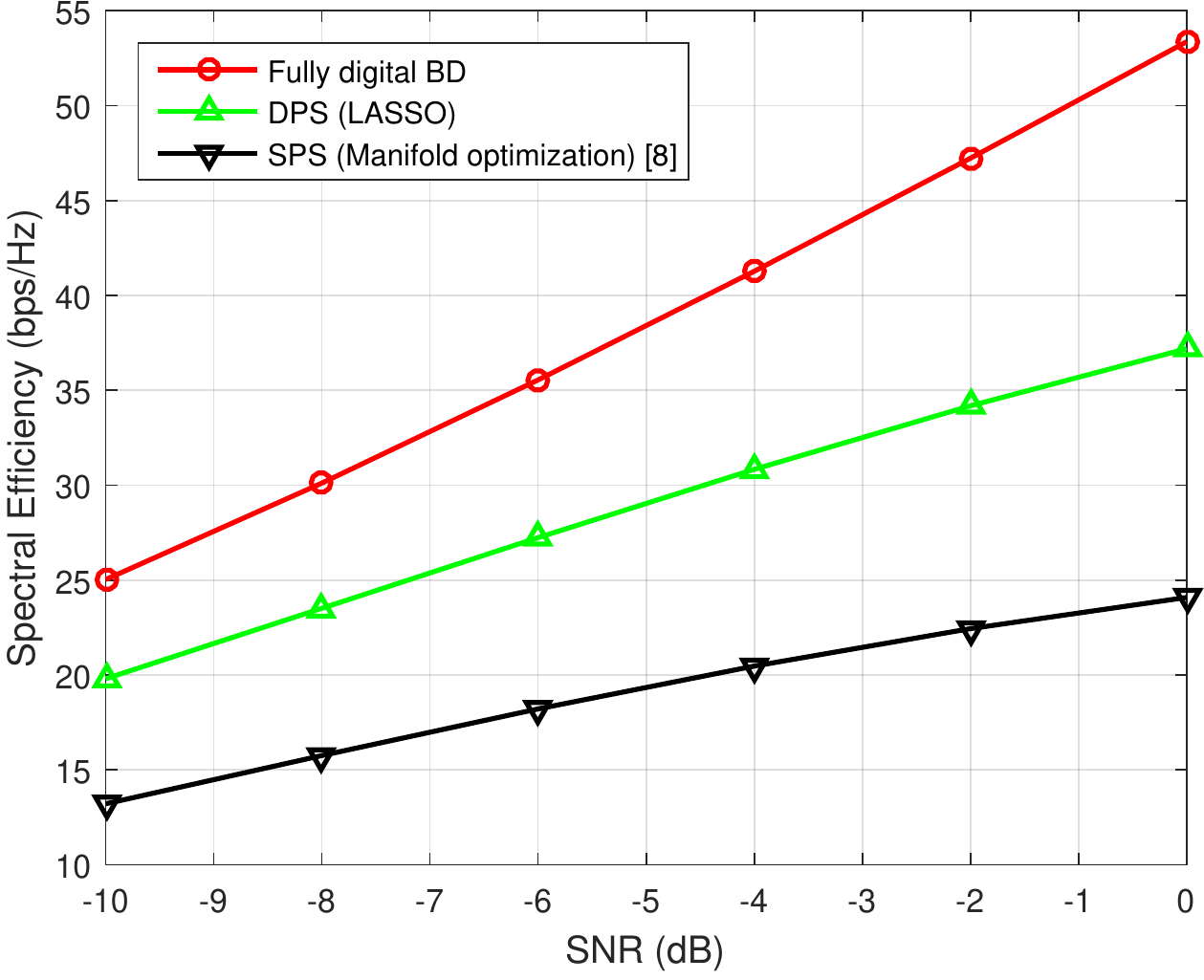}
		\caption{Spectral efficiency achieved by different hybrid precoder implementations when $\Nt=64$, $\Nr=9$, $K=5$, $N_s=2$, and $\NRFt=K\NRFr=KN_s$.}\label{fig3}
	\end{figure}
	
	Fig. \ref{fig3} shows the spectral efficiency of two different hybrid precoder implementations. First, we see that the proposed DPS implementation outperforms the conventional SPS implementation. This performance gain is obtained from doubling the phase shifters in the analog RF precoder since we adopt the same digital precoder for both implementations. More importantly, the LASSO formulation of the RF-only precoding in Section \ref{SecIIIA} enables low-complexity algorithms and results in a closed-form solution when the digital precoder is semi-orthogonal. In contrast, the algorithm based on manifold optimization involves a complicated iterative procedure and hence leads to high computational complexity \cite{7397861}. With the conventional SPS implementation, there always exists an obvious trade-off between the design complexity and performance \cite{7397861,6928432,7389996}. On the contrary, with the DPS implementation, by modifying the hardware implementation, we can significantly reduce the design complexity and improve the system performance in the meanwhile.
	
	\subsection{Hybrid Precoding in the Fully-Connected Structure}
\begin{figure}[tbp]
		\begin{minipage}[t]{0.49\linewidth}
			\centering\includegraphics[height=5.75cm]{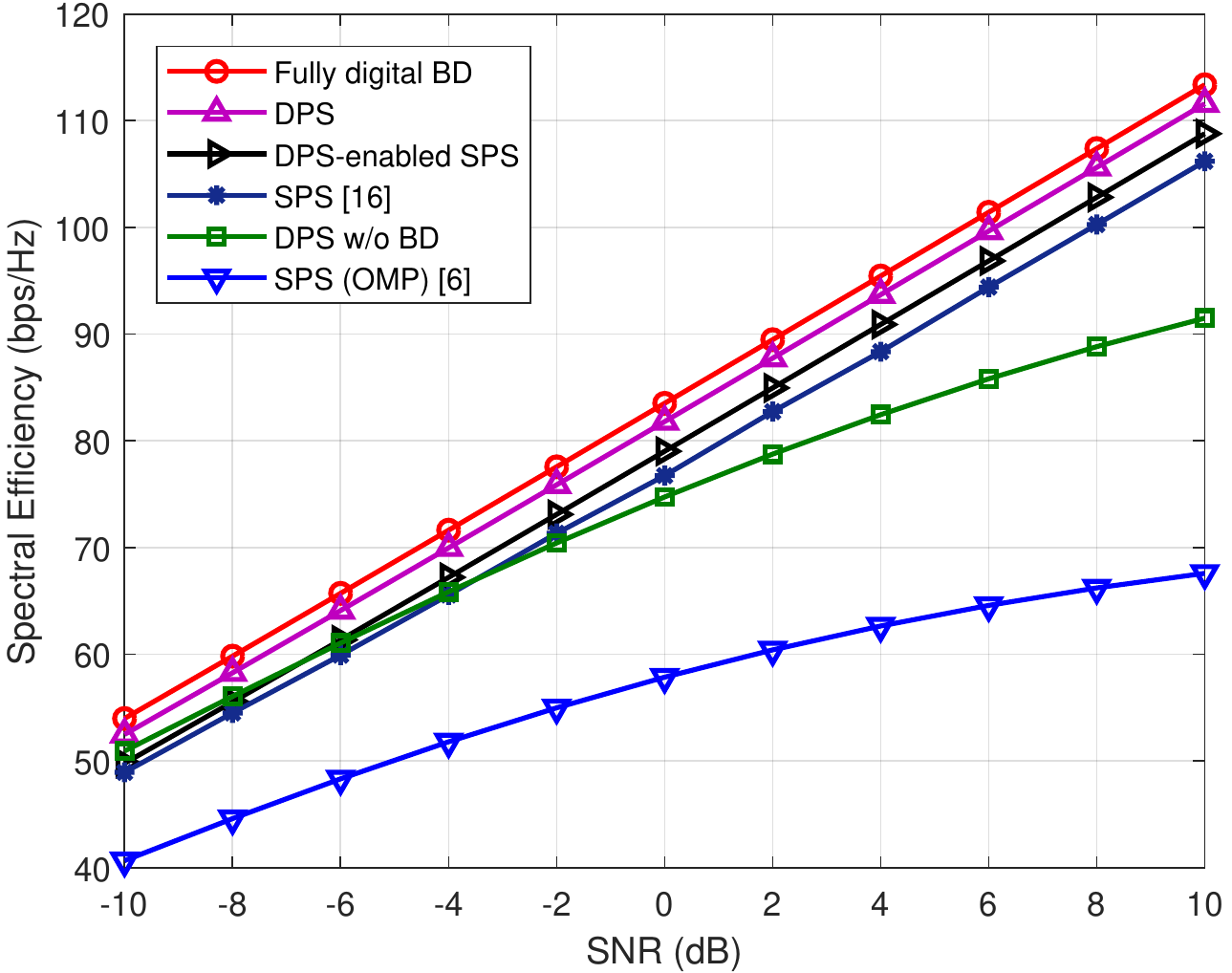}
			\caption{Spectral efficiency achieved by different hybrid precoding algorithms in the fully-connected structure when $\Nt=256$, $\Nr=16$, $K=3$, and $N_s=3$.}\label{fig1}
		\end{minipage}
		\begin{minipage}[t]{0.49\linewidth}
			\centering\includegraphics[height=5.75cm]{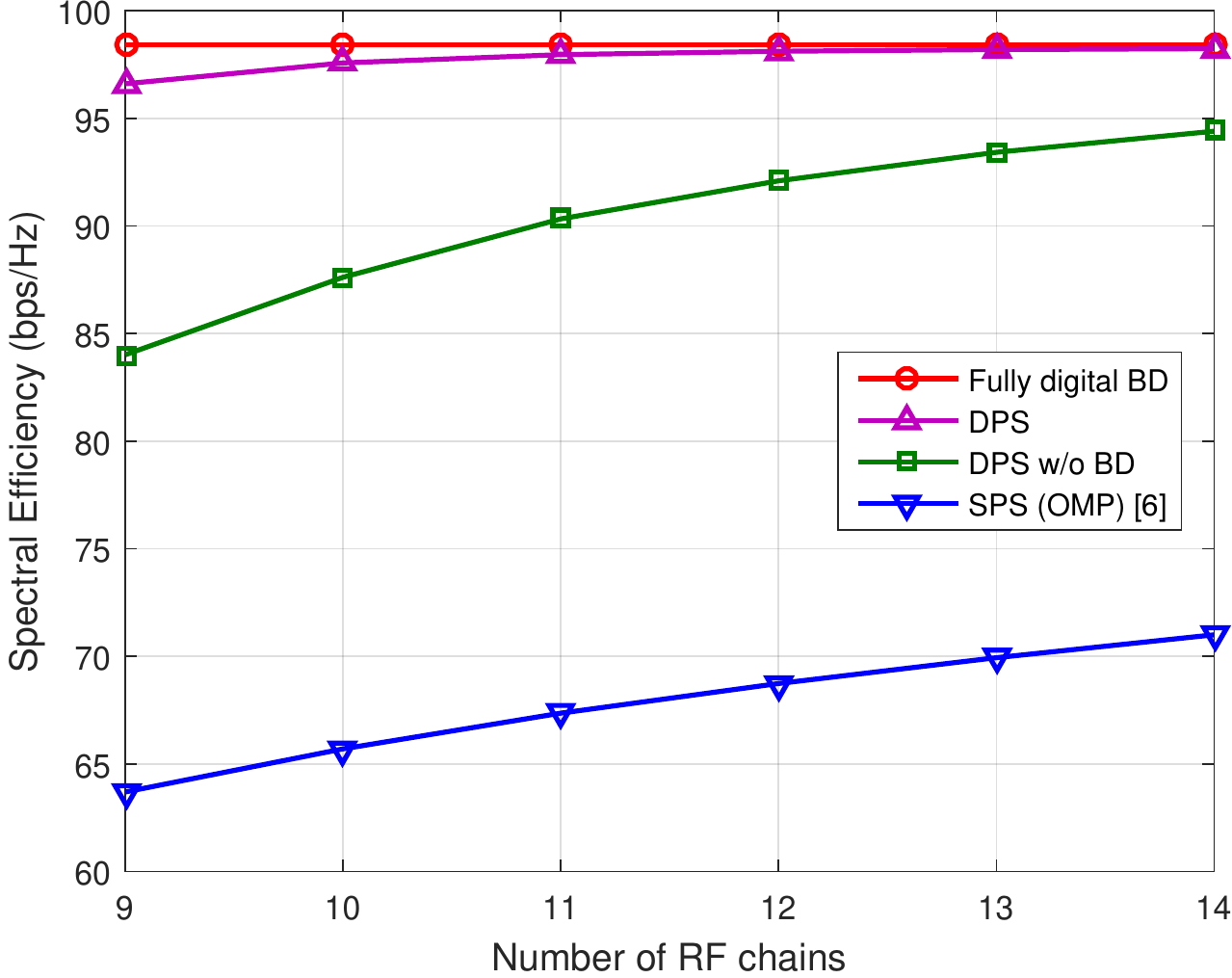}
			\caption{Spectral efficiency achieved by different hybrid precoding algorithms in the fully-connected structure for different transmit RF chain numbers $\NRFt$, given $\Nt=256$, $\Nr=16$, $K=3$, $N_s=3$, $\NRFr=N_s$, and SNR=5 dB.}\label{fig2}
		\end{minipage}
	\end{figure}

	Fig. \ref{fig1} shows the spectral efficiency achieved by different algorithms with the minimum numbers of RF chains, i.e., $\NRFt=KN_s$ and $\NRFr=N_s$. 
	{\color{black}To illustrate the effectiveness of the proposed implementation and algorithms, two hybrid precoding algorithms with the SPS implementation \cite{6717211,7913599} are adopted  as benchmarks.	
	The widely used OMP algorithm \cite{6717211} is with the lowest computational complexity among all the available algorithms with the conventional SPS implementation. 
	The algorithm in \cite{7913599} achieves the best spectral efficiency in the literature for multiuser OFDM systems with the SPS implementation, which iteratively optimizes the phases in the analog precoder,  therefore with  high design complexity.}
	
	First, we see that the DPS implementation outperforms the existing algorithm in \cite{7913599} with the conventional SPS implementation, and approaches the fully digital one.
	This means that the proposed algorithm can more accurately approximate the fully digital precoder than existing algorithms, even though the RF chains are limited.
	While the OMP is already with the lowest computational complexity among all the available algorithms for the conventional SPS implementation, the proposed design approach with simple matrix decomposition enjoys even lower computational complexity and much higher spectral efficiency than the OMP algorithm. This shows that the proposed DPS implementation achieves both good spectral efficiency and computational efficiency.
	
	In \cite{7397861,7037444,7389996}, it has been pointed out that approximating the fully digital precoder with a hybrid structure will lead to a near optimal performance in single-user single-carrier, single-user multicarrier, and multiuser single-carrier mm-wave MIMO systems, respectively. Next, we will verify whether this insight still holds in more general multiuser OFDM mm-wave MIMO systems.
	In Fig. \ref{fig1}, we evaluate the performance of the proposed hybrid precoding without the additional BD operation mentioned in Section \ref{SecIIIC}, i.e., only approximating the fully digital precoder. We discover that, without the BD precoder canceling the interuser interference, there will be residual interuser interference, which results in an obvious performance loss compared to the fully digital one, especially at high SNRs. This phenomenon illustrates that simply approximating the fully digital precoder with the hybrid one is not sufficient in multiuser multicarrier mm-wave systems. Although interuser interference also exists in multiuser single-carrier systems, this issue is more prominent in the multicarrier system as the analog precoder is shared by a large number of subcarriers. The comparison in Fig. \ref{fig1} demonstrates the effectiveness and necessity of the additional BD operation proposed in Section \ref{SecIIIC}. 
	
	In addition, the performance of the DPS-enabled SPS design is presented. Fig. \ref{fig1} shows that it only entails little performance loss compared to the DPS one, and outperforms the spectral efficiency of the state-of-the-art algorithm in \cite{7913599}. This phenomenon empirically verifies the effectiveness of the  phase extraction operation discussed in Section \ref{SecIIID}. Note that, with the SPS implementation, existing design algorithms  involve iterative procedures \cite{6717211,7913599}. In contrast, the proposed DPS-enabled one offers a low-complexity design with closed-form solutions. 
	{\color{black}Overall, the proposed DPS-enabled SPS approach provides an effective heuristic way to design SPS hybrid precoders, which enjoy both lower computational complexity and higher achievable spectral efficiency than existing algorithms.}
	
	Fig. \ref{fig2} compares different precoding schemes for different RF chain numbers $\NRFt$ at the BS side while keeping $\NRFr=N_s$ as the minimum number of RF chains at each user.
	It is shown that the DPS implementation  approaches the performance of the fully digital precoder when the number of RF chains is slightly larger than the number of transmitted data streams, which cannot be realized by the existing OMP algorithm. There is no need to further increase the number of RF chains considering the increased power consumption and hardware complexity.
	
	\begin{figure}[tbp]
		\centering
		\includegraphics[height=5.75cm]{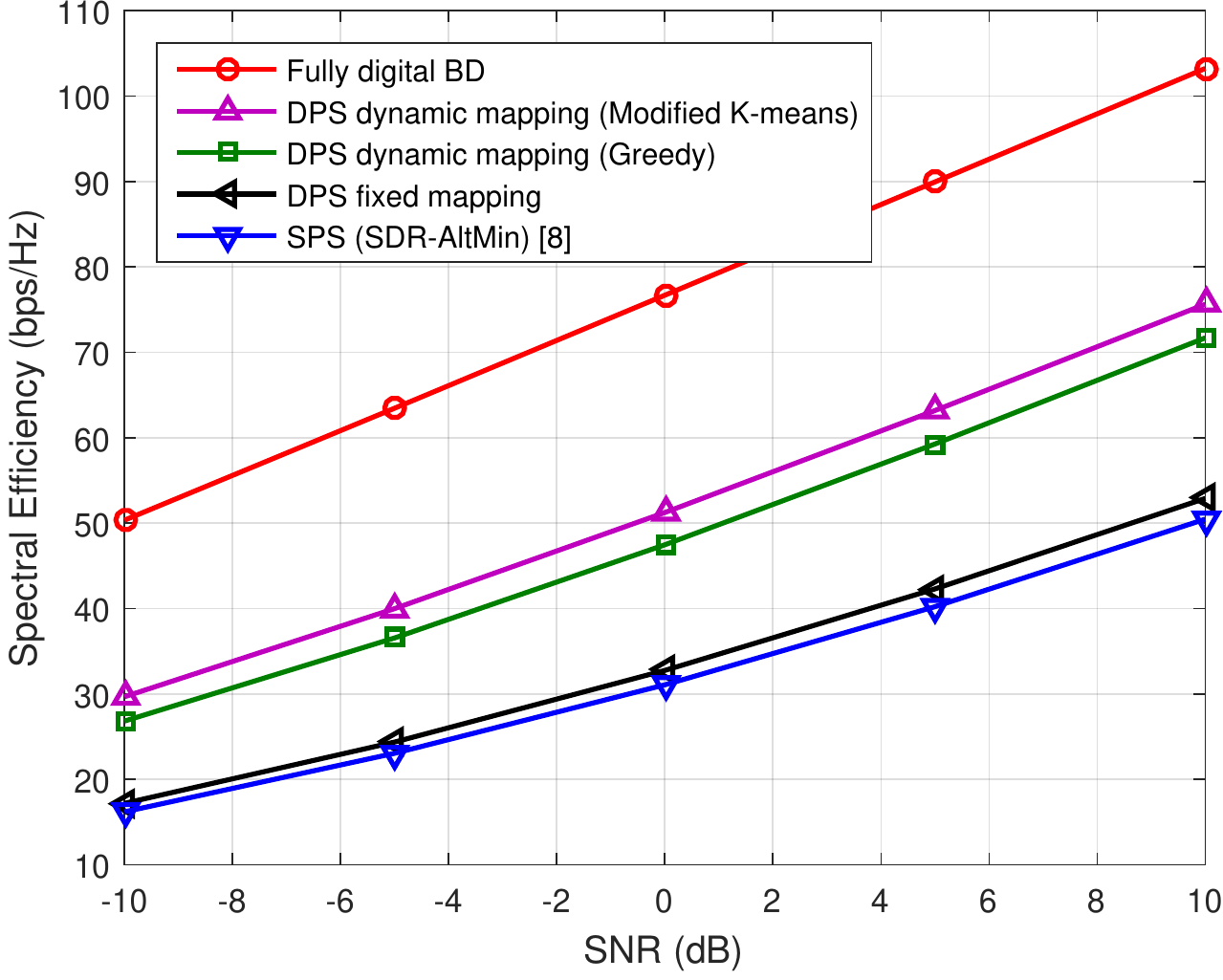}
		\caption{Spectral efficiency achieved by different hybrid precoding algorithms in the partially-connected structure when $\Nt=256$, $\Nr=16$, $K=4$, and $N_s=2$.}\label{fig4}
	\end{figure}
	\subsection{Hybrid Precoding in the Partially-Connected Structure}
	Fig. \ref{fig4} shows the performance of the proposed design approaches in the partially-connected structure with the minimum numbers of RF chains, i.e., $\NRFt=KN_s$ and $\NRFr=N_s$.
	We see that, due to the sharply reduced number of phase shifters, the partially-connected structure does entail non-negligible performance loss compared to the fully digital one. Furthermore, although the computational complexity of the proposed algorithms is significantly reduced compared to the SDR-AltMin algorithm with the SPS implementation in \cite{7397861}, simply doubling the number of phase shifters with the fixed mapping only has little performance gain over the conventional SPS implementation. However, Fig. \ref{fig4} demonstrates that dynamic mapping is able to shrink the gap between the fixed mapping and the fully digital precoding by half.

	Furthermore, we compare the performance of the proposed greedy and modified K-means algorithms for the dynamic mapping in the partially-connected structure. Fig. \ref{fig4} illustrates that, although the modified K-means algorithms only converges to a local optimum of \eqref{eq34}, it achieves a higher spectral efficiency than the greedy algorithm. Even though the cluster number $\NRFt$ is much smaller than the number of observation vectors $\Nt$ in the modified K-means algorithm, it turns out that the algorithm converges quickly in the simulation\footnote{The modified K-means algorithm converges within 10	iterations for almost all the channel realizations.}. Overall, the modified K-means algorithm steps up as an excellent dynamic mapping design algorithm for hybrid precoding in the partially-connected structure.
	
	\section{Conclusions}\label{SecVI}
	This paper proposed a new DPS implementation for hybrid precoding, based on which computationally efficient algorithms were developed for different hybrid precoder structures in general multiuser OFDM mm-wave systems. 
	The comparisons between different hybrid precoder implementations and structures are listed in Table I, which help reveal important design guidelines as follows.
	\begin{table}[tbp]
		\centering
		\caption{Comparisons between different hybrid precoder implementations and structures}
		\begin{tabular}{|c|c|c|c|c|c|}\hline
			\textbf{Imple-} & \multirow{2}[0]{*}{\textbf{Structure}} & \multirow{2}[0]{*}{\textbf{Design approach}}  & \textbf{Hardware complexity}& \textbf{Computational} & \multirow{2}[0]{*}{\textbf{Performance}} \\
			\textbf{mentation} &       &  &    \textbf{(No. of phase shifters)}   & \textbf{complexity} &  \\
			\hhline{|=|=|=|=|=|=|}
			\multirow{2}[0]{*}{SPS} & Fully-connected &   MO-AltMin \cite{7397861}    & $\NRFt\Nt$   &  Extremely high    &  \checkmark\checkmark\checkmark\\\cline{2-6}
			& Partially-connected & SDR-AltMin \cite{7397861}     & $\Nt$  &  High     &\checkmark  \\\hline
			\multirow{2}[0]{*}{DPS} & Fully-connected & Matrix decomposition      & $2\NRFt(\Nt-\NRFt)$  &  $\mathcal{O}\left(\NRFt^2\Nt F\right)$    &\checkmark\checkmark\checkmark\checkmark  \\\cline{2-6}
			& Partially-connected & Modified K-means      &   $2\Nt$   &  $\mathcal{O}\left(N\NRFt^2\Nt F\right)$    &\checkmark\checkmark  \\\hline
		\end{tabular}%
		\begin{tablenotes}
			\item * For each structure, we pick the design approach with the best performance among existing works. More \checkmark means higher spectral efficiency. The comparisons of the computational complexity between the SPS and DPS implementations are based on the running time in the simulation. 
			The computational complexity for the DPS implementation is calculated with the minimum number of RF chains, i.e., $\NRFt=KN_s$.
		\end{tablenotes}
	\end{table}%
	\begin{itemize}
		\item For the DPS fully-connected structure, the computational complexity of the hybrid precoder design is significantly reduced, with the performance approaching the fully digital one when the number of RF chains is comparable to that of the data streams. For cost and power consideration, it is unnecessary to further employ more RF chains.
		\item Dynamic mapping is shown to be critical to improve the spectral efficiency in the DPS partially-connected structure, which is basically a clustering problem. Although there exists some performance loss compared with the fully-connected one, due to the much diminished hardware complexity, the partially-connected structure serves as an economic choice for hybrid precoding.
		\item Overall, it is beneficial, from both performance and computational complexity points of view, to adopt the DPS implementation in hybrid precoding, in comparison to the conventional SPS implementation. Interestingly, the fully-connected structure enjoys lower computational complexity than the partially-connected one, as dynamic mapping brings additional complexity in order to enhance the performance.
	\end{itemize}
	{In summary, the proposed DPS implementation enjoys advantages of high spectral efficiency and high computational efficiency, at the cost of more hardware components. We envision that once low-cost high-resolution commercial phase shifters are available, or for cost-insensitive applications, the DPS implementation would be an ideal choice. On the other hand, our investigation on this new implementation also provides valuable guidelines on designing other hybrid precoder structures, as summarized in Remark 1.} While this paper demonstrated the great potential of the proposed DPS implementation, more works will be needed to investigate other problems involving hybrid precoder design, e.g., to consider the hybrid precoder design combined with channel training and feedback.
{\color{black}	Furthermore, more  efforts are needed on quantitative comparisons between different hybrid precoder structures considering realistic hardware components.}

	\appendices
	\section{Proof of Lemma \ref{lem2}}\label{appA}
	Since both \eqref{analogp} and \eqref{lasso} are convex problems, it is equivalent to prove that the dual problem of \eqref{lasso} is \eqref{analogp}. We introduce a dummy variable $\mathbf{y}$ to derive the dual as follows.
	\begin{equation}\label{eq36}
	\begin{aligned}
	&\underset{\mathbf{x},\mathbf{y}}{\mathrm{minimize}} && \frac{1}{2}||\mathbf{Ax-b}||_2^2+2||\mathbf{y}||_1\\
	&\mathrm{subject\thinspace to}&&\mathbf{y=x}.
	\end{aligned}
	\end{equation} 
	The Lagrangian of \eqref{eq36} is
	\begin{equation}
	\mathcal{L}(\mathbf{x},\mathbf{y};\mathbf{v})=\frac{1}{2}||\mathbf{Ax-b}||_2^2+2||\mathbf{y}||_1+\mathbf{v}^H(\mathbf{x-y}).
	\end{equation}
	Therefore, the dual function is given by
	\begin{equation}\label{eq38}
	\underset{\mathbf{v}}{\mathrm{maximize}}\thickspace\underset{\mathbf{x},\mathbf{y}}{\inf}\quad\frac{1}{2}||\mathbf{Ax-b}||_2^2+\mathbf{v}^H\mathbf{x}+2||\mathbf{y}||_1-\mathbf{v}^H\mathbf{y}.
	\end{equation}
	By the definition of the conjugate function to the $\ell_1$-norm, we can obtain that
	\begin{equation}\label{eq39}
	\underset{\mathbf{y}}{\inf}\quad2||\mathbf{y}||_1-\mathbf{v}^H\mathbf{y}=-\underset{\mathbf{y}}{\sup}\quad\mathbf{v}^H\mathbf{y}-2||\mathbf{y}||_1=\begin{cases}
	0&\text{if }||\mathbf{v}||_\infty\le2\\
	-\infty&\text{otherwise}.
	\end{cases}
	\end{equation}
	By checking the first order optimality condition of $\mathbf{x}$, we can get
	\begin{equation}\label{eq40}
	\frac{\partial}{\partial\mathbf{x}}\mathcal{L}(\mathbf{x},\mathbf{y};\mathbf{v})=\mathbf{A}^H(\mathbf{Ax-b})+\mathbf{v}=\mathbf{0}\Rightarrow\mathbf{x}=\left(\mathbf{A}^H\mathbf{A}\right)^{-1}\left(\mathbf{A}^H\mathbf{b-v}\right).
	\end{equation}
	{By substituting \eqref{eq39} and \eqref{eq40} into \eqref{eq38}, we can derive the objective function as follows:
	\begin{equation}\label{eq47}
	\underset{||\mathbf{v}||_\infty\le2}{\mathrm{maximize}}\thickspace\frac{1}{2}||\mathbf{A}\left(\mathbf{A}^H\mathbf{A}\right)^{-1}\left(\mathbf{A}^H\mathbf{b-v}\right)-\mathbf{b}||_2^2+\mathbf{v}^H\left(\mathbf{A}^H\mathbf{A}\right)^{-1}\left(\mathbf{A}^H\mathbf{b-v}\right).
	\end{equation}
	Expand \eqref{eq47} and drop the constant terms, and then we can conclude the dual problem as follows}
	\begin{equation}\label{eq41}
	\underset{||\mathbf{v}||_\infty\le2}{\mathrm{minimize}}\quad\frac{1}{2}\mathbf{v}^H\left(\mathbf{A}^H\mathbf{A}\right)^{-1}\mathbf{v}-\Re\{\mathbf{v}^H\left(\mathbf{A}^H\mathbf{A}\right)^{-1}\mathbf{A}^H\mathbf{b}\}.
	\end{equation}
	Multiply the objective function in \eqref{eq41} by $2$ and add a constant term $\mathbf{b}^H\mathbf{A}\left(\mathbf{A}^H\mathbf{A}\right)^{-2}\mathbf{A}^H\mathbf{b}$, we can recast the dual problem as
	\begin{equation}
	\begin{aligned}
	&\underset{\mathbf{v}}{\mathrm{minimize}} && ||\mathbf{f-Dv}||_2^2\\
	&\mathrm{subject\thinspace to}&&|\mathbf{v}(i)|\le2.
	\end{aligned}
	\end{equation} 
	where $\mathbf{D}^H\mathbf{D}=\left(\mathbf{A}^H\mathbf{A}\right)^{-1}$ and $\left(\mathbf{A}^H\mathbf{A}\right)^{-1}\mathbf{A}^H\mathbf{b}=\mathbf{D}^H\mathbf{f}$.
	By defining $\mathbf{f}=\fopt$, $\mathbf{v}=\fRF$, $\mathbf{D}^H\mathbf{D}$, and $\mathbf{D}=\mathbf{F}_\mathrm{BB}^T\otimes \mathbf{I}_{\Nt}$, and inversely applying the matrix vectorization, it completes the proof of Lemma \ref{lem2}.

	\ifCLASSOPTIONcaptionsoff
	\newpage
	\fi
	
	\bibliographystyle{IEEEtran}
	\bibliography{bare_jrnl}
	
\end{document}